\documentclass{article}
\usepackage[latin9]{inputenc}
\usepackage[a4paper]{geometry}
\usepackage{amsmath}
\usepackage{graphicx}
\usepackage{url}
\makeatletter

\usepackage{authblk}

\providecommand{\keywords}[1]{
	\small	
	\textbf{\textit{Keywords---}} #1
}

\author[a,b]{Zedong Bi}
\author[b,c,d,e,1]{Changsong Zhou} 

\affil[a]{Institute for Future, Qingdao University, Shandong 266071, China}
\affil[b]{Department of Physics, Centre for Nonlinear Studies and Institute of Computational and Theoretical Studies, Hong Kong Baptist University, Kowloon Tong, Hong Kong, China}

\affil[c]{Research Centre, HKBU Institute of Research and Continuing Education, Shenzhen, China}

\affil[d]{Beijing Computational Science Research Center, Beijing, China}
\affil[e]{Department of Physics, Zhejiang University, 38 Zheda Road, Hangzhou, China}
\affil[1]{E-mail: cszhou@hkbu.edu.hk}

\makeatother

\begin{document}
	
\title{Understanding the computation of time using neural network models }
	
\maketitle

\section*{Abstract}
	To maximize future rewards in this ever-changing world, animals must be able to discover the temporal structure of stimuli and then anticipate or act correctly at the right time. How the animals perceive, maintain, and use time intervals ranging from hundreds of milliseconds to multi-seconds in working memory? How temporal information is processed concurrently with spatial information and decision making? Why there are strong neuronal temporal signals in tasks in which temporal information is not required? A systematic understanding of the underlying neural mechanisms is still lacking. Here, we addressed these problems using supervised training of recurrent neural network models. We revealed that neural networks perceive elapsed time through state evolution along stereotypical trajectory, maintain time intervals in working memory in the monotonic increase or decrease of the firing rates of interval-tuned neurons, and compare or produce time intervals by scaling state evolution speed. Temporal and non-temporal information are coded in subspaces orthogonal with each other, and the state trajectories with time at different non-temporal information are quasi-parallel and isomorphic. Such coding geometry facilitates the decoding generalizability of temporal and non-temporal information across each other. The network structure exhibits multiple feedforward sequences that mutually excite or inhibit depending on whether their preferences of non-temporal information are similar or not. We identified four factors that facilitate strong temporal signals in non-timing tasks, including the anticipation of coming events. Our work discloses fundamental computational principles of temporal processing, and is supported by and gives predictions to a number of experimental phenomena.

\keywords{interval timing $|$ population coding $|$ neural network model} 

\section*{Significance}
Perceiving, maintaining, and using time intervals in working memory are crucial for animals to anticipate or act correctly at the right time in the ever-changing world. Here we systematically study the underlying neural mechanisms by training recurrent neural networks to perform temporal tasks or complex tasks in combination with spatial information processing and decision making. We found that neural networks perceive time through state evolution along stereotypical trajectories, and produce time intervals by scaling evolution speed. Temporal and non-temporal information are jointly coded in a way that facilitates decoding generalizability. We also provided potential sources for the temporal signals observed in non-timing tasks. Our study revealed the computational principles of a number of experimental phenomena and provided several novel predictions.

\section*{Introduction}
Much information that the brain processes and stores is temporal
in nature. Therefore, to understand the processing of time in the
brain is of fundamental importance in neuroscience \cite{Merchant_2013,Allman_2014,Paton_2018,Petter_2018}.
To predict and maximize future rewards in this ever-changing world,
animals must be able to discover the temporal structure of stimuli
and then flexibly anticipate or act correctly at the right time.
To this end, animals must be able to perceive, maintain, and then
use time intervals in working memory, appropriately combining the
processing of time with spatial information and decision making.
Based on behavioral data and the diversity of neuronal response profiles,
it has been proposed \cite{Mauk_2004,Buhusi_2005} that time intervals
in the range of hundreds of milliseconds to multi-seconds can be
decoded through neuronal population states evolving along transient
trajectories. The neural mechanisms may be accumulating firing \cite{Treisman_1963,Killeen_1988},
synfire chains \cite{Hahnloser_2002,Lynch_2016}, the beating of
a range of oscillation frequencies \cite{Matell_2004}, etc. However,
these mechanisms are challenged by recent finding that animals can
flexibly adjust the evolution speed of population activity along
an invariant trajectory to produce different intervals \cite{Wang_2018}.
Through behavioral experiments, it was found that humans can store
time intervals as distinct items in working memory in a resource
allocation strategy \cite{Teki_2014}, but an electrophysiological
study on the neuronal coding of time intervals maintained in working
memory is still lacking. Moreover, increasing evidence indicates
that timing does not rely on dedicated circuits in the brain, but
instead is an intrinsic computation that emerges from the inherent
dynamics of neural circuits \cite{Ivry_2008,Paton_2018}. Spatial
working memory and decision making are believed to rely mostly on
a prefronto-parietal circuit \cite{Lara_2015,Gold_2007}. The dynamics
and the network structure that enable this circuit to combine spatial
working memory and decision making with flexible timing remains unclear.
Overall, our understanding of the processing of time intervals in
the brain is fragmentary and incomplete. It is therefore essential
to develop a systematic understanding of the fundamental principle
of temporal processing and its combination with spatial information
processing and decision making.

\begin{figure}[tbph]
\center \includegraphics[scale=0.6]{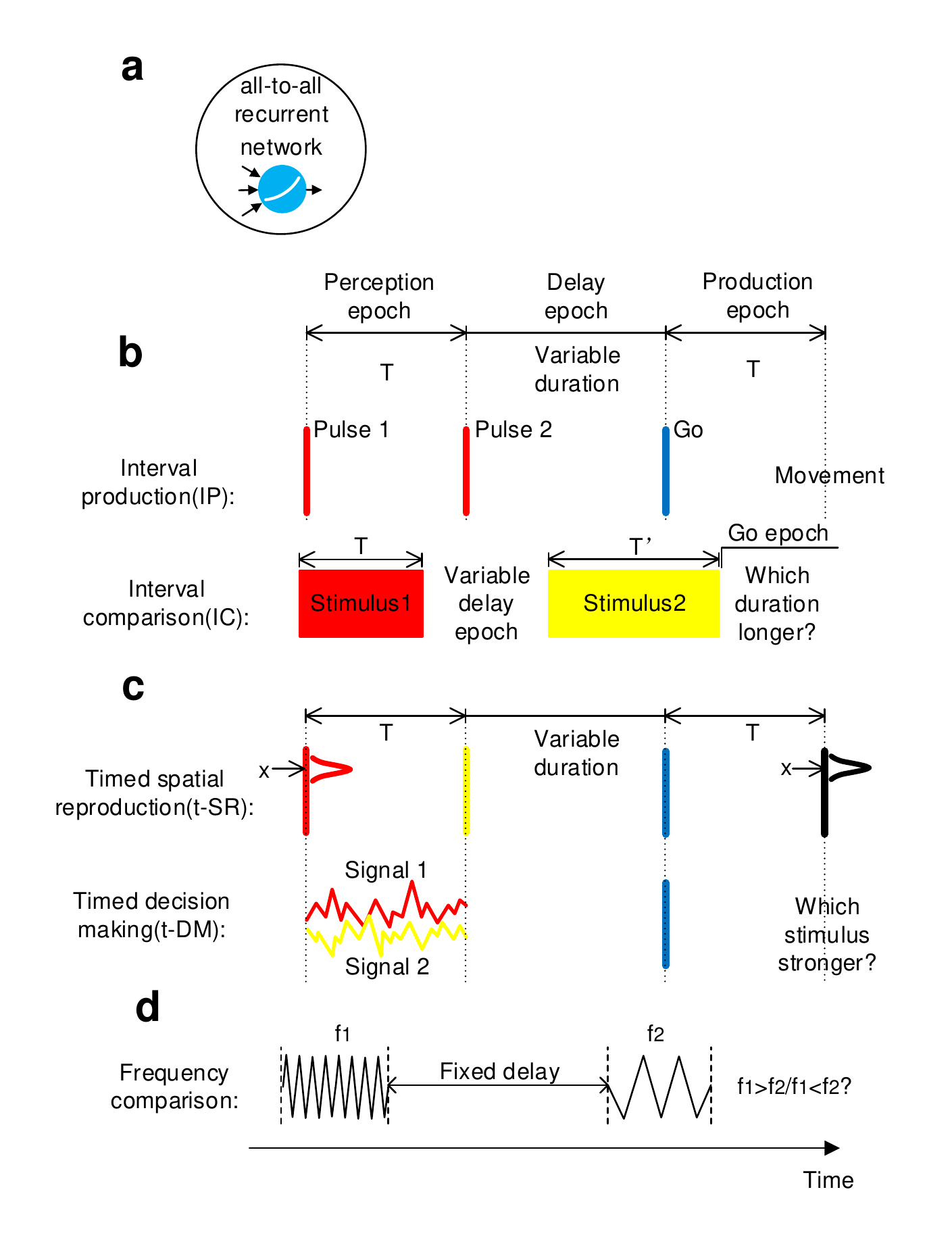}

\protect\protect\caption{\textbf{\label{fig:Model-setup}Model setup.} (\textbf{a}) All-to-all
connected recurrent networks with softplus units are trained. (\textbf{b})
Basic timing tasks. IP: The duration $T$ of the perception epoch
determines the movement time after the Go cue. IC: The duration $T$
of the stimulus1 epoch is compared with the duration $T'$ of the
stimulus2 epoch. Stimuli with different colors (red, yellow, or blue)
indicate that they are input to the network through different synaptic
weights. (\textbf{c}) Combined timing tasks. $T$ determines the
movement time after the Go cue. Spatial location (t-SR) or decision
choice (t-DM) determines the movement behavior. (\textbf{d}) A non-timing
task in the experimental study \cite{Machens_2010}. Although the
duration of the delay period is not needed to perform the task, there
exists strong temporal signals in the delay period. }
\end{figure}

The formation of temporal signals in the brain is another unexplored
question. Strong temporal signals were found in the brain even when
monkeys performed working memory tasks where temporal information
was not needed \cite{Romo_1999,Machens_2010,Kobak_2016,Murray_2017,Rossi-Poola_2019}.
In a vibrotactile working memory task \cite{Romo_1999}, monkeys
were trained to report which of the two vibrotactile stimuli separated
by a fixed-delay period had higher frequency (\textbf{Fig. \ref{fig:Model-setup}d}).
Surprisingly, although the duration of the delay period was not needed
to perform this task, temporal information was still coded in the
neuronal population state during the delay period, with the time-dependent
variance explaining more than 75\% of the total variance \cite{Machens_2010,Kobak_2016}.
Similar scenario was also found in other non-timing working memory
tasks \cite{Kobak_2016,Murray_2017,Rossi-Poola_2019}. It is unclear
why so strong temporal signals arised in non-timing tasks.

Previous works showed that after being trained to perform tasks such
as categorization, working memory, decision making, and motion generation,
artifical neural networks (ANN) exhibited coding or dynamic properties
surprisingly similar to experimental observations \cite{Hong_2016,Song_2016,Mante_2013,Sussillo_2015}.
Compared with animal experiments, ANN can cheaply and easily implement
a series of tasks, greatly facilitating the test of various hypotheses
and the capture of common underlying computational principles \cite{Yang_2019,Orhan_2019}.
In this paper, we trained recurrent neural networks (\textbf{Fig.
\ref{fig:Model-setup}a}) to study the processing of temporal information.
Firstly, by training networks on basic timing tasks which require
only temporal information to perform (\textbf{Fig. \ref{fig:Model-setup}b}),
we studied how time intervals are perceived, maintained, and used
in working memory. Secondly, by training networks on combined timing
tasks which require both temporal and non-temporal information to
perform (\textbf{Fig. \ref{fig:Model-setup}c}), we studied how the
processing of time is combined with spatial information processing
and decision making, the influence of this combination to decoding
generalizability, and the network structure that this combination
is based on. Thirdly, by training networks on non-timing tasks (\textbf{Fig.
\ref{fig:Model-setup}d}), we studied why so large time-dependent
variance arises in non-timing tasks, thereby understanding the factors
that facilitate the formation of temporal signals in the brain. Our
work presents a thorough understanding of the neural computation
of time.

\section*{Results}

We trained a recurrent neural network (RNN) of 256 softplus units
supervisedly using back-propagation through time. Self-connections
of the RNN were initialized to 1, and off-diagonal connections were
initialized as independent Gaussian variables with mean 0 \cite{Orhan_2019}, with different training configurations initialized using different random seeds.
The strong self-connections supported self-sustained activity after
training (\textbf{Fig. S1b}), and the non-zero initialization of
the off-diagonal connections induced sequential activity comparable
to experimental observations \cite{Orhan_2019}. We stopped training
as soon as the performance of the network reached criterion \cite{Song_2016,Sussillo_2015}
(see performance examples in \textbf{Fig. S1}).

\subsection*{Basic timing tasks: interval production and interval comparison
tasks}

\subsubsection*{Interval production task}

\begin{figure*}
\centering \includegraphics[scale=0.6]{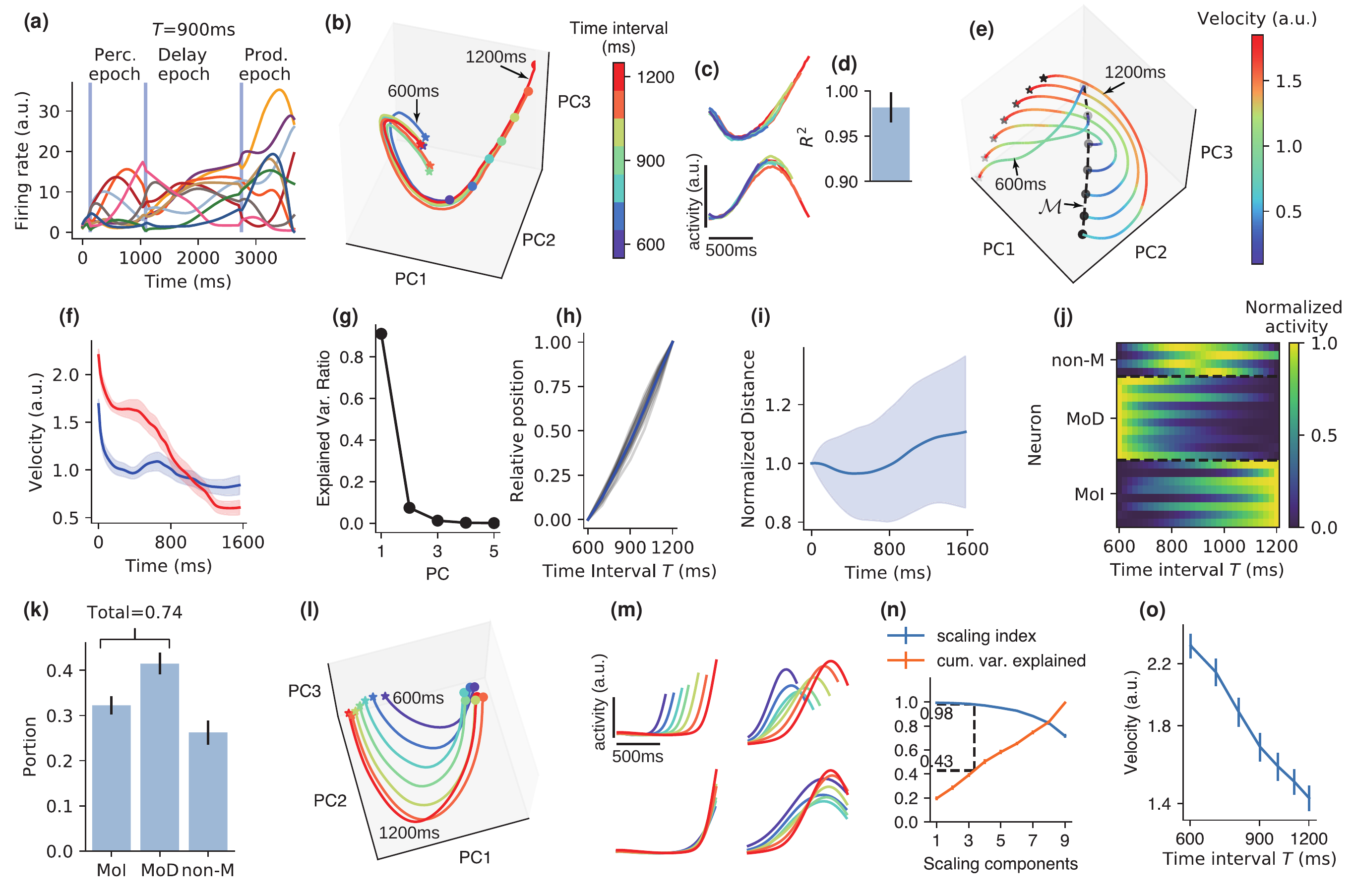}

\protect\protect\caption{\textbf{\label{fig:IP_task}Interval production task.} (\textbf{a})
The activities of example neurons (indicated by lines of different
colors) when the time interval $T$ between the first two pulses
is 900 ms. Vertical blue shadings indicate the pulses input to the network. (\textbf{b}) Population
activity in the perception epoch in the subspace of the first three
PCs. Colors indicate the time interval $T$. Stars and circles respectively
indicate the starting and ending points of the perception epoch.
The trajectories for $T=600$ ms and 1200 ms are labeled. (\textbf{c})
Firing profiles of two example neurons in the perception epoch. Line
colors have the same meaning as in panel \textbf{b}. (\textbf{d})
Coefficient of determination ($R^{2}$) of how much the neuronal
firing profile with the largest $T$ can explain the variance of
the firing profiles with smaller $T$ in the perception epoch. Error
bar indicates s.d. over different neurons and $T$ values. (\textbf{e})
Population activity in the subspace of the first three PCs in the
delay epoch. Colors indicate trajectory speed. The increasing blackness
of stars and circles indicates trajectories with $T=600$ ms, 700
ms,$\cdots$,1200 ms. The dashed curve connecting the end points
of the delay epoch marks manifold $\mathcal{M}$. (\textbf{f}) Trajectory
speed as a function of time in the delay epoch when $T=600$ ms (blue)
and 1200 ms (red). Shaded belts indicate s.e.m. (standard error of
mean) over training configurations. (\textbf{g}) Ratio of explained
variance of the first five PCs of manifold $\mathcal{M}$. Error
bars that indicate s.e.m. are smaller than plot markers. (\textbf{h})
The position of the state at the end of the delay epoch projected in the first
PC of manifold $\mathcal{M}$ as a function of $T$. The position
when $T=600$ ms (or 1200 ms) is normalized to be 0 (or 1). Gray
curves: 16 training configurations. Blue curve: mean value. (\textbf{i})
The distance between two adjacent curves in the delay epoch as a
function of time, with the distance at the beginning of the delay
epoch normalized to be 1. Shaded belts indicate standard deviation.
(\textbf{j}) Firing rates of example neurons of monotonically decreasing
(MoD), monotonically increasing (MoI), and non-monotonic (non-M)
types as functions of $T$ in manifold $\mathcal{M}$. (\textbf{k})
The portions of the three types of neurons. (\textbf{i}) Population
activity in the production epoch in the subspace of the first three
PCs. Colors indicate the time intervals to be produced, as shown
in the color bar of panel \textbf{b}. Stars and circles respectively
indicate the starting and ending points of the production epoch.
(\textbf{m}) Upper: firing profiles of two example neurons in the
production epoch. Lower: firing profiles of the two neurons after
temporally scaled according to produced intervals. (\textbf{n}) A
point at horizontal coordinate $x$ means the scaling index (blue)
or ratio of explained variance (orange) of the subspace spanned by
the first $x$ scaling components (SCs). Dashed lines indicate that
a subspace with scaling index 0.98 explains, on average, 43\% of
the total variance. (\textbf{o}) Trajectory speed in the subspace
of the first three SCs as the function of the time interval to be
produced. In panels \textbf{k}, \textbf{n}, \textbf{o}, error bars
indicate s.e.m. over training configurations.  During training, we
added recurrent and input noises (see Methods). Here and in the following,
when analyzing the network properties after training, we turned off
noises by default. We kept noises for the perception epoch in panels\textbf{
b},\textbf{ c},\textbf{ d}. Without noise, the trajectories in the
perception epoch would fully overlap under different $T$s.}
\end{figure*}

In the interval production (IP) task (the first task of \textbf{Fig.
\ref{fig:Model-setup}b}), the network was to perceive the interval
$T$ between the first two pulses, maintain the interval during the
delay epoch with variable duration, and then produce an action at
time $T$ after the Go cue. Neuronal activities after training exhibited
strong fluctuations (\textbf{Fig. \ref{fig:IP_task}a}). In the following,
we report on the dynamics of the network in the perception, delay
and production epochs of IP (see \textbf{Fig.\ref{fig:Model-setup}
}for illustration of these epochs).

The first epoch is the perception epoch. In response to the first
stimulus pulse, the network started to evolve from almost the same
state along an almost identical trajectory in different simulation
trials with different $T$ values until another pulse came (\textbf{Fig. \ref{fig:IP_task}b});
the activities of individual neurons before the second pulse in different
trials highly overlapped (\textbf{Fig. \ref{fig:IP_task}c, d}). Therefore, the network state evolved along a stereotypical trajectory starting from the first pulse,
and the time interval $T$ between the first two pulses can be read out using the position in this trajectory when the second pulse came. Behaviorally, a human's
perception of the time interval between two acoustic pulses is impaired
if a distractor pulse appears shortly before the first pulse \cite{Karmarkar_2007}.
A modeling work \cite{Karmarkar_2007} explained that this is because
successful perception requires the network state to start to evolve
from near a state $\mathbf{s}_{0}$ in response to the first pulse,
whereas the distractor pulse kicks the network state far away from
$\mathbf{s}_{0}$. This explanation is consistent with our results
that interval perception requires a stereotypical trajectory.

We then studied how the information of timing interval $T$ between
the first two pulses was maintained during the delay epoch. We have
the following findings. (1) The speeds of the trajectories decreased
with time in the delay epoch (\textbf{Fig. \ref{fig:IP_task}e, f}).
(2) The states $\mathbf{s}_{\text{EndDelay}}$ at the end of the
delay epoch at different $T$s were aligned in a manifold $\mathcal{M}$
whose first PC explained 90\% of its variance (\textbf{Fig. \ref{fig:IP_task}g}).
(3) For a specific simulation trial, the position of $\mathbf{s}_{\text{EndDelay}}$
in manifold $\mathcal{M}$ linearly encoded the $T$ value of the
trial (\textbf{Fig. \ref{fig:IP_task}h}). (4) The distance between
two adjacent trajectories kept almost unchanged with time during
the delay, neither decayed to zero, nor exploded (\textbf{Fig. \ref{fig:IP_task}i}):
this stable dynamics supported the information of $T$ encoded by
the position in the stereotypical trajectory at the end of the perception
epoch in being maintained during the delay. Collectively, $\mathcal{M}$
approximated a line attractor \cite{Seung_1996,Mante_2013} with
slow dynamics, and $T$ was encoded as the position in $\mathcal{M}$.
To better understand the scheme of coding $T$ in $\mathcal{M}$,
we classified neuronal activity $f(T)$ in manifold $\mathcal{M}$
as a function of $T$ into three types (\textbf{Fig. \ref{fig:IP_task}j,
k}): monotonically decreasing (MoD), monotonically increasing (MoI),
and non-monotonic (non-M) (see Methods). We found that most neurons
were MoD or MoI, whereas only a small portion were non-M neurons
(\textbf{Fig. \ref{fig:IP_task}k}). This implies that the network
mainly used a complementary (i.e., concurrently increasing and decreasing)
monotonic scheme to code time intervals in the delay epoch, similar
to the scheme revealed in Ref. \cite{Mita_2009,Romo_1999}. This
dominance of monotonic neurons may be the reason why the first PC
of $\mathcal{M}$ explained so much variance (\textbf{Fig. \ref{fig:IP_task}g}),
see Section S2 and \textbf{Figs. S2g, h} for a simple explanation.

In the production epoch, the trajectories of the different $T$ values
tended to be isomorphic (\textbf{Fig. \ref{fig:IP_task}l}). The
neuronal activity profiles were self-similar when stretched or compressed
in accordance with the produced interval (\textbf{Fig. \ref{fig:IP_task}m}),
suggesting temporal scaling with $T$ \cite{Wang_2018}. To quantify
this temporal scaling, we defined the scaling index (SI) of a subspace
$\mathcal{S}$ as the portion of variance of the projections of trajectories
into $\mathcal{S}$ that can be explained by temporal scaling \cite{Wang_2018}.
We found that the distribution of SI of individual neurons aggregated
toward 1 (\textbf{Fig. S2b}), and the first two PCs that explained
most variance have the highest SI (\textbf{Fig. S2c}). We then used
a dimensionality reduction technique that furnished a set of orthogonal
directions (called scaling components, or SCs) in the network state
space that were ordered according to their SI (see Methods). We found
that a subspace (spanned by the first three SCs) that had high SI
(=0.98) occupied about 40\% of the total variance of trajectories
(\textbf{Fig. \ref{fig:IP_task}n}), in contrast with the low SI of the perception epoch (\textbf{Fig. S2f}). The average speed of the trajectory
in the subspace of the first three SCs was inversely proportional
to $T$ (\textbf{Fig. \ref{fig:IP_task}o}). Collectively, the network
adjusted its dynamic speed to produce different time intervals in
the production epoch, similar to observations of the medial frontal
cortex of monkeys \cite{Remington_2018,Wang_2018}. Additionally,
we found a non-scaling subspace whose mean activity during the production
epoch changed linearly with $T$ (\textbf{Fig. S2d, e}), also similar
to the experimental observations in Ref. \cite{Remington_2018,Wang_2018}.

\subsubsection*{Interval comparison task}

In the interval comparison (IC) task (the second task of\textbf{
Fig. \ref{fig:Model-setup}b}), the network was successively presented
two intervals; it was then required to judge which interval was longer.
IC required the network to perceive the time interval $T$ of the
stimulus1 epoch, to maintain the interval in the delay epoch, and
to use it in the stimulus2 epoch whose duration is $T'$. Similar
to IP, the network perceived time interval with a stereotypical trajectory
in the stimulus1 epoch (\textbf{Fig. S3a-c}) and maintained time
interval using attractor dynamics with a complementary monotonic
coding scheme in the delay epoch (\textbf{Fig. S3d-h}). The trajectory
in the stimulus2 epoch had a critical point $\mathbf{s}_{crit}$
at time $T$ after the start of stimulus 2. The network was to give
different comparison outputs at the Go epoch depending on whether
or not the trajectory had passed $\mathbf{s}_{crit}$ at the end
of stimulus 2. To make a correct comparison choice, only the period
from the start of stimulus 2 to $\mathbf{s}_{crit}$ (or to the end
of stimulus 2 if $T>T'$) need to be timed: as long as the trajectory
had passed $\mathbf{s}_{crit}$, the network could readily make the
decision that $T<T'$, with no more timing required. After training,
we studied the trajectories from the start of stimulus 2 to $\mathbf{s}_{crit}$
in the cases that $T<T'$, and found temporal scaling (\textbf{Fig.
S3j-n}) similar to the production epoch of IP, consistently with animal
experiments \cite{Leon_2003,Mendoza_2018}. These similarities between
IP and IC on how to perceive, maintain and use time intervals imply
universal computational schemes for neural networks to process temporal
information.

The average speed of the trajectory after $\mathbf{s}_{crit}$ increased
with $T$ (\textbf{Fig. S3o}), whereas the speed before $\mathbf{s}_{crit}$
decreased with $T$: this implies that the dynamics after $\mathbf{s}_{crit}$
was indeed different from that before. 

\begin{figure*}[th]
	\center \includegraphics[scale=0.6]{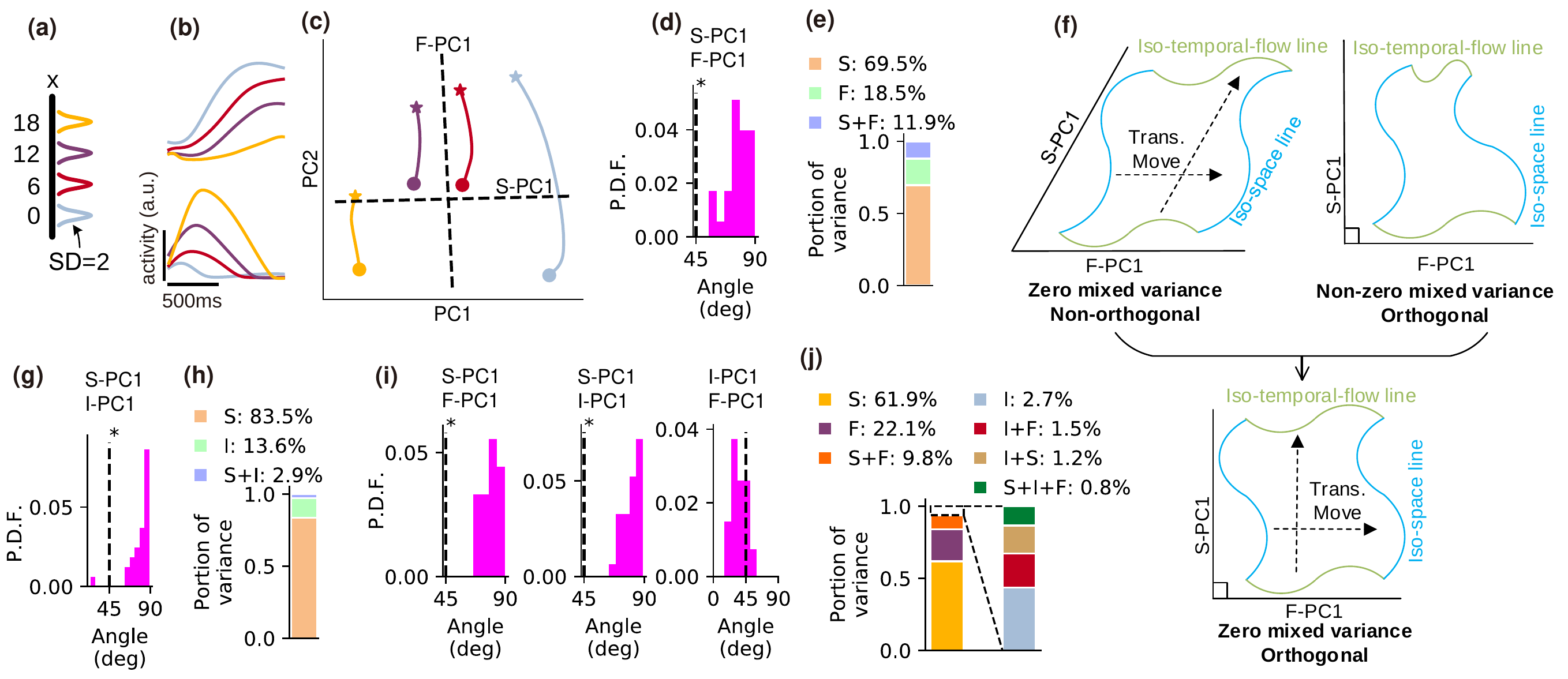}
	
	\protect\protect\protect\caption{\textbf{\label{fig:t-SR}Timed spatial reproduction task.} (\textbf{a})
		Color scheme that represents the spatial location of the first pulse,
		used in panels \textbf{b, c}. Location is represented by Gaussian
		bump with standard deviation 2. (\textbf{b}) Firing profiles of four
		example neurons in the perception epoch. (\textbf{c}) Trajectory
		of the perception epoch in the subspace of the first two PCs. Stars
		indicate the points after 400 ms of transient period from the beginning
		of the perception epoch, and circles indicate the ending points of
		the perception epoch. Dashed lines represent the projections of F-PC1
		and S-PC1 in this subspace. (\textbf{d}) Probability distribution
		function (p.d.f.) of the angle between F-PC1 and S-PC1 in the perception
		epoch over 32 training configurations. Asterisk indicates significant
		($p<0.05$) larger than $45^{\circ}$ (t test). (\textbf{e}) Portion
		of variance explained by spatial information (S), temporal flow (F)
		and their mixture (S+F) in the perception epoch, averaging over 32 training configurations. (\textbf{f}) Schematic
		for the meanings of angle and mixed variance. Zero mixed variance
		implies that different iso-space (blue) or iso-temporal-flow (green)
		lines are related by translational movement, forming parallelogram-like
		grids (upper left), together with orthogonality (upper right) implies
		rectangle-like grids (lower). (\textbf{g}) The distribution of the
		angle between I-PC1 and S-PC1 in manifold $\mathcal{M}$ (\textbf{Fig.\ref{fig:IP_task}e})
		at the end of the delay epoch. (\textbf{h}) The portion of variance
		explained by spatial information (S), time interval (I) and their
		mixture (S+I) in manifold $\mathcal{M}$ at the end of the delay
		epoch. (\textbf{i}) The distributions of the angles between F-PC1,
		I-PC1 and S-PC1 in the production epoch. (\textbf{j}) The portion
		of variance explained by spatial information (S), temporal flow (F),
		time interval (I) and their mixtures in the production epoch. In
		\textbf{a-e}, $T=1200$ ms for the perception epoch; in \textbf{g-j},
		$T=600$ ms,700 ms,$\cdots$,1200 ms for the delay and production
		epochs. }
\end{figure*}

\subsection*{Combined timing tasks: timed spatial reproduction and timed decision
making tasks}

It is a ubiquitous phenomenon that neural networks encode more than one quantities simultaneously \cite{Ashe_1994,Churchland_2012,Lankarany_2019}. In this subsection, we will discuss how neural networks encode temporal and spatial information (or decision choice) simultaneously, which enables the brain to take the right action at the right time. 

\subsubsection*{Timed spatial reproduction task}

In the timed spatial reproduction (t-SR) task (the first task in
\textbf{Fig. \ref{fig:Model-setup}c}), the network was to not only
take action at the desired time but also act at the spatial location
indicated by the first pulse. Similar to IP and IC, the network used
stereotypical trajectories, attractors and speed scaling to perceive,
maintain and produce time intervals (\textbf{Fig. S4}). In the following,
we will focus on the coding combination of temporal and spatial information.

In the perception epoch, under the two cases when the first pulse
were at two locations $x$ and $y$ separately, the activities $r_{i,\text{perc}}(t,x)$
and $r_{i,\text{perc}}(t,y)$ of the $i$th neuron exhibited similar
profiles with time $t$ (\textbf{Fig. \ref{fig:t-SR}b}), especially
when $x$ and $y$ had close values. In our simulation, the location
of the first pulse was represented by a Gaussian bump with standard
deviation $2$, which is much smaller than the smallest spatial distance
$6$ between two different colors in \textbf{Fig. \ref{fig:t-SR}a,
b}; thus, the similarity of the temporal profiles in \textbf{Fig.
\ref{fig:t-SR}b} should not result from the overlap of the sensory
inputs from the first pulse but rather emerge during training. 

To quantitatively investigated the coding combination of temporal
and spatial information, we studied the first temporal-flow PC (F-PC1)
of the neuronal population, namely the first PC of $\{\langle r_{i,\text{perc}}(t,x)\rangle_{x}\}_{i}$,
and the first spatial PC (S-PC1), namely the first PC of $\{\langle r_{i,\text{perc}}(t,x)\rangle_{t}\}_{i}$,
with $\langle\cdot\rangle_{a}$ indicating averaging over parameter
$a$. By \textit{temporal flow}, we mean the time elapsed from the
beginning of a specific epoch. We found that the angle between F-PC1
and S-PC1 distributed around $90^{\circ}$, significantly larger
than $45^{\circ}$ (\textbf{Fig. \ref{fig:t-SR}d}). This indicates
that temporal flow and spatial information was coded in almost orthogonal
subspaces (\textbf{Fig. \ref{fig:t-SR}c}). We then studied the mixed
variance \cite{Kobak_2016}. Specifically, the variance explained
by temporal (or spatial) information is $v_{t}=\text{Var}_{i,t}(\langle r_{i,\text{perc}}(t,x)\rangle_{x})$
(or $v_{x}=\text{Var}_{i,x}(\langle r_{i,\text{perc}}(t,x)\rangle_{t})$),
and the mixed variance is $v_{t+x}=v_{tot}-v_{t}-v_{x}$, where $v_{tot}=\text{Var}_{i,t,x}(r_{i,\text{perc}}(t,x))$
is the total variance. We found that the mixed variance took a small
portion of the total variance, smaller than the variance of either
temporal or spatial information (\textbf{Fig. \ref{fig:t-SR}e}).
To understand the implication of this result, we noted that a sufficient
condition for $v_{t+x}=0$ is that different iso-space (or iso-temporal-flow)
lines are related with each other through translational movement
(\textbf{Fig. \ref{fig:t-SR}f, upper left}), where an iso-space
(or iso-temporal-flow) line is a manifold in the state space with
different temporal flow (or space) values but a fixed space (or temporal
flow) value; the opposite extreme case $v_{t+x}=v_{tot}$ implies
that different iso-space (or iso-temporal-flow) lines are strongly
intertwined, see SI Text Section S3 for details. Together, orthogonality
and small mixed variance suggest that iso-space and iso-time lines
interweave into rectangle-like grids (\textbf{Fig. \ref{fig:t-SR}f,
lower}), see \textbf{Fig. S6} for illustrations of the simulation
results. 

In the delay epoch, the population states were attracted toward a
manifold $\mathcal{M}$ of slow dynamics at the end of the delay
epoch (\textbf{Fig. \ref{fig:IP_task}e-i} and \textbf{Fig. S4b-d}), maintaining
both the duration $T$ of the perception epoch and the spatial information
$x$. We studied the coding combination of $T$ and $x$ in $\mathcal{M}$
in a similar way to above. We found that the first time-interval
PC (I-PC1), namely the first PC to code $T$, was largely orthogonal
with the first spatial PC (S-PC1) (\textbf{Fig. \ref{fig:t-SR}g}),
and the mixed variance between $T$ and $x$ was small (\textbf{Fig.\ref{fig:t-SR}h}). 

\begin{figure}
	\center \includegraphics[scale=0.7]{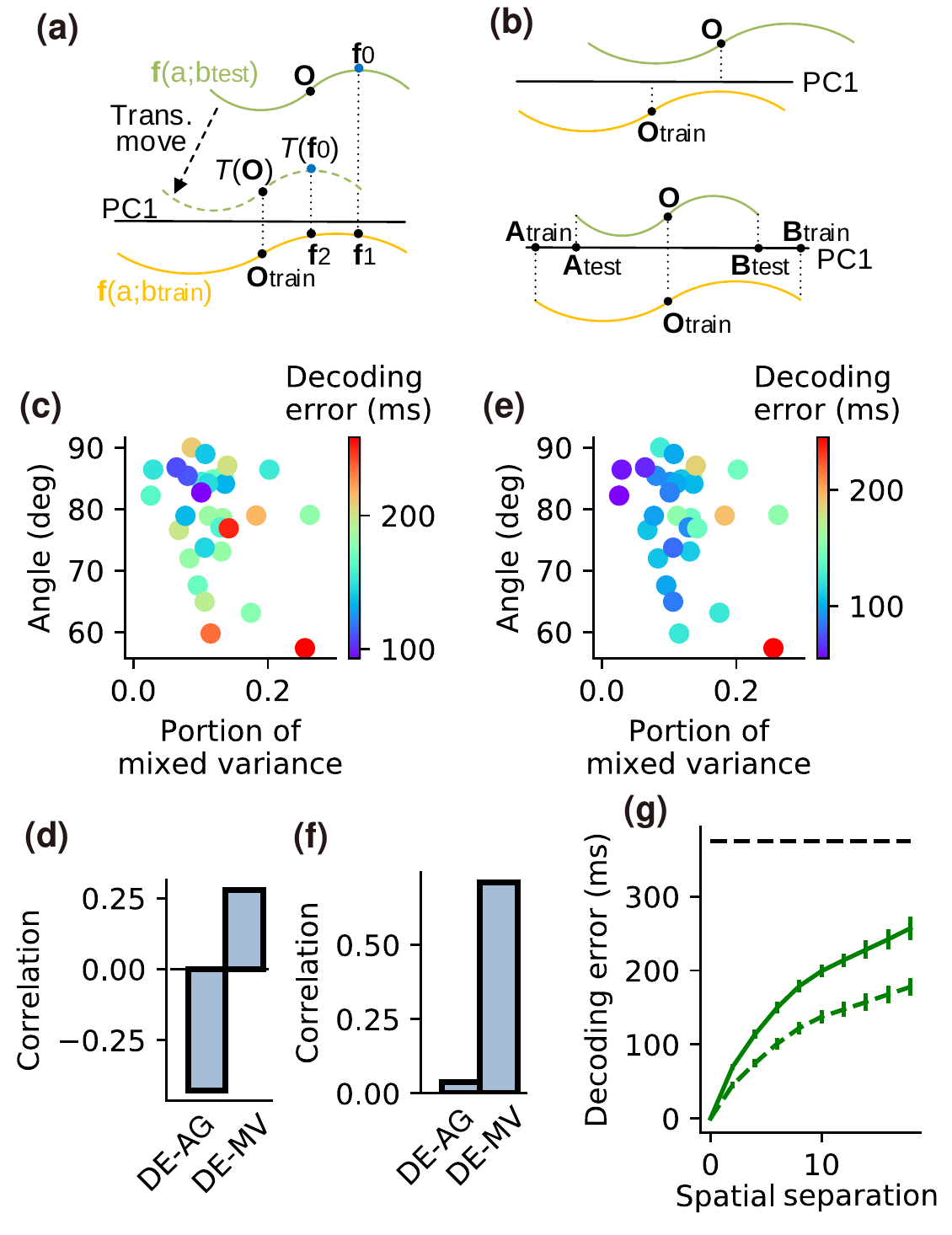}
	
	\protect\protect\protect\caption{\label{fig:Decoding-generalizability}\textbf{Decoding generalizability.}
		(\textbf{a}) Schematic that explains Decoder 1 and Decoder 2. The
		decoders read the value of $a$ through state $\mathbf{f}_{0}$ in
		iso-$b$ line $\mathbf{f}(a;b_{test})$ (green) after being trained
		by another iso-$b$ line $\mathbf{f}(a;b_{train})$ (orange). Decoder
		1 reads $a$ to be the same as that of $\mathbf{f}_{1}$, because
		$\mathbf{f}_{0}$ and $\mathbf{f}_{1}$ project to the same point
		on PC1 (black horizontal line) of $\mathbf{f}(a;b_{train})$. Decoder
		2 first translationally moves $\mathbf{f}(a;b_{test})$ so that its
		mass center $\mathcal{T}(O)$ after translational movement $\mathcal{T}$
		projects to the same point as the mass center $O_{train}$ of $\mathbf{f}(a;b_{train})$
		on PC1, and then reads $a$ according to $\mathcal{T}(\mathbf{f}_{0})$,
		which is the $a$ value of $\mathbf{f}_{2}$. (\textbf{b}) Two error
		sources of Decoder 1. Upper: the mass centers $O$ and $O_{train}$
		do not project to the same point on PC1. Lower: the projections of
		$\mathbf{f}(a;b_{train})$ and $\mathbf{f}(a;b_{test})$ on PC1 (lines
		$A_{train}B_{train}$ and $A_{test}B_{test}$) do not have the same
		length. (\textbf{c}) The error of Decoder 1 (indicated by dot color)
		to read temporal flow across different spatial locations as a function
		of the angle and mixed variance between the temporal-flow and spatial
		subspaces, in the production epoch of t-SR task. (\textbf{d}) Correlation
		between decoding error (DE) and angle (AG), and between DE and mixed
		variance (MV). (\textbf{e},\textbf{f}) The same as \textbf{c }and
		\textbf{d}, except for Decoder 2. (\textbf{g}) Decoding error as
		a function of $|x_{train}-x_{test}|$, after Decoder 1 (solid line)
		or Decoder 2 (dashed line) is trained to read the temporal flow using
		the iso-space line at spatial location $x_{train}$, and then tested
		at spatial location $x_{test}$. Horizontal dashed line indicates
		chance level, supposing the decorder works by random guess. Error
		bars represent mean$\pm$s.e.m. across simulation trials. Panels
		\textbf{c-f }analyze the data averaging over $|x_{train}-x_{test}|$
		in individual training configurations. $T=1200$ ms. See decoding
		generalizability in other epochs of t-SR task and t-DM task in \textbf{Figs.
			S7, S8}.}
\end{figure}

In the production epoch, the network needed to maintain three information:
temporal flow $t$, time interval $T$, and spatial location $x$.
We studied the angle between the first PCs of any two of them (i.e.,
F-PC1, I-PC1 and S-PC1). We found that S-PC1 was orthogonal with
F-PC1 and I-PC1, but F-PC1 and I-PC1 was not orthogonal (\textbf{Fig.
\ref{fig:t-SR}i}). For any two parameters, their mixed variance
was smaller than the variance of their own (\textbf{Fig. \ref{fig:t-SR}j}),
see Methods for details.

Collectively, in all the three epochs, the coding subspaces of temporal
and spatial information were largely orthogonal with small mixed
variance, suggesting rectangle-like grids of iso-space and iso-time
lines, see \textbf{Fig. S6} for illustrations.

\subsubsection*{Timed decision making task}

In the timed decision making (t-DM) task (the second task in \textbf{Fig.
\ref{fig:Model-setup}c}), the network was to make a decision choice
at the desired time to indicate which of the two presented stimuli
was stronger. Similar to IP, IC and t-SR, the network used stereotypical
trajectories, attractors and speed scaling to separately perceive,
maintain and produce time intervals (\textbf{Fig. S5a-m}). In all
the three epochs of t-DM, the first PC to code decision choice (D-PC1)
was orthogonal with F-PC1 or I-PC1, and the mixed variance between
any two parameters were small (\textbf{Fig. S5n-s}); but F-PC1 and
I-PC1 in the production epoch was not orthogonal (\textbf{Fig. S5r}).
These results are all similar to those of t-SR task.

\subsection*{Decoding generalizability}

We then studied how the above geometry of coding space influences
decoding generalizability: suppose the population state space is
parameterized by $a$ and $b$, we want to know the error of decoding
$a$ from a state $\mathbf{f}_{0}$ in an iso-$b$ line $\mathbf{f}(a;b_{test})$
after training the decoder using another iso-$b$ line $\mathbf{f}(a;b_{train})$
(\textbf{Fig. \ref{fig:Decoding-generalizability}a}). We considered
two types of nearest-centroid decoders \cite{Murray_2017}: Decoder
1 projects both $\mathbf{f}_{0}$ and $\mathbf{f}(a;b_{train})$
into the first PC of $\mathbf{f}(a;b_{train})$, and reads the value
of $a$ to be the value that minimized the distance between $\mathcal{P}_{dec}[\mathbf{f}(a;b_{train})]$
and $\mathcal{P}_{dec}[\mathbf{f}_{0}]$, where $\mathcal{P}_{dec}[\cdot]$
indicates the projection operation; Decoder 2 first translationally
moves the whole iso-$b$ line $\mathbf{f}(a;b_{test})$ so that the
mass center of $\mathcal{P}_{dec}[\mathcal{T}[\mathbf{f}(a;b_{test})]]$
coincides with that of $\mathcal{P}_{dec}[\mathbf{f}(a;b_{train})]$,
where $\mathcal{T}$ indicates the translation operation, and then
reads $a$ according to $\mathcal{P}_{dec}[\mathcal{T}[\mathbf{f}_{0}]]$
(\textbf{Fig. \ref{fig:Decoding-generalizability}a}). Apparently,
zero error of Decoder 1 requires $\mathcal{P}_{dec}[\mathbf{f}(a;b_{test})]$
and $\mathcal{P}_{dec}[\mathbf{f}(a;b_{train})]$ to perfectly overlap.
If the grids woven by iso-$a$ and iso-$b$ lines are tilted (\textbf{Fig.
\ref{fig:t-SR}f, upper left}) or non-parallelogram-like (\textbf{Fig.\ref{fig:t-SR}f,
upper right}), which can be respectively quantified by the orthogonality
or mixed variance ratio introduced in the above section, the projections
$\mathcal{P}_{dec}[\mathbf{f}(a;b_{test})]$ and $\mathcal{P}_{dec}[\mathbf{f}(a;b_{train})]$
may have non-overlapping mass centers (\textbf{Fig. \ref{fig:Decoding-generalizability}b,
upper}) or different lengths (\textbf{Fig. \ref{fig:Decoding-generalizability}b,
lower}), causing decoding error. Decoder 2 translationally moves
the mass center of $\mathcal{P}_{dec}[\mathbf{f}(a;b_{test})]$ to
the position of that of $\mathcal{P}_{dec}[\mathbf{f}(a;b_{train})]$,
so its decoding error only depends on the non-parallelogram-likeness
of grids. Biologically, the projection onto the first PC of $\mathbf{f}(a;b_{train})$
can be realized by Hebbian learning of decoding weights \cite{Dayan_2001},
the nearest-centroid scheme can be realized by winner-take-all decision
making \cite{Murray_2017}, and the overlap of the mass centers in
Decoder 2 can be realized by homeostatic mechanisms \cite{Turrigiano_2011} to keep the mean neuronal activity over different iso-$b$ lines unchanged (\textbf{eq. S19}). 

Consistently with the decoding scenario above, when decoding temporal
flow generalizing across spatial information in the production epoch
of t-SR, the error of Decoder 1 negatively correlated with the angle
$\theta$ between the first temporal-flow PC and the first spatial
PC, and positively correlated with the portion $\rho_{mix}$ of mixed
variance (\textbf{Fig. \ref{fig:Decoding-generalizability}c, d});
whereas the error of Decoder 2 depended weakly on $\theta$, and
positively correlated with $\rho_{mix}$ (\textbf{Fig. \ref{fig:Decoding-generalizability}e,
f}), see Methods for details. Thanks to the angle orthogonality and
small mixed variance (\textbf{Fig. \ref{fig:t-SR}}), both decoders have above-chance performance
(\textbf{Fig. \ref{fig:Decoding-generalizability}g}). Additionally,
for both t-SR and t-DM tasks, we studied the decoding generalization
of temporal (non-temporal) information across non-temporal (temporal)
information in all the perception, delay and production epochs. In all
cases, we found how the decoding error depended on the angle between
the first PCs of the decoded and generalized variables and the mixed
variance followed similar scenario to above (\textbf{Figs. S7, S8}).

\subsection*{Sequential activity and network structure}

\begin{figure}[t]
	\center\includegraphics[scale=0.7]{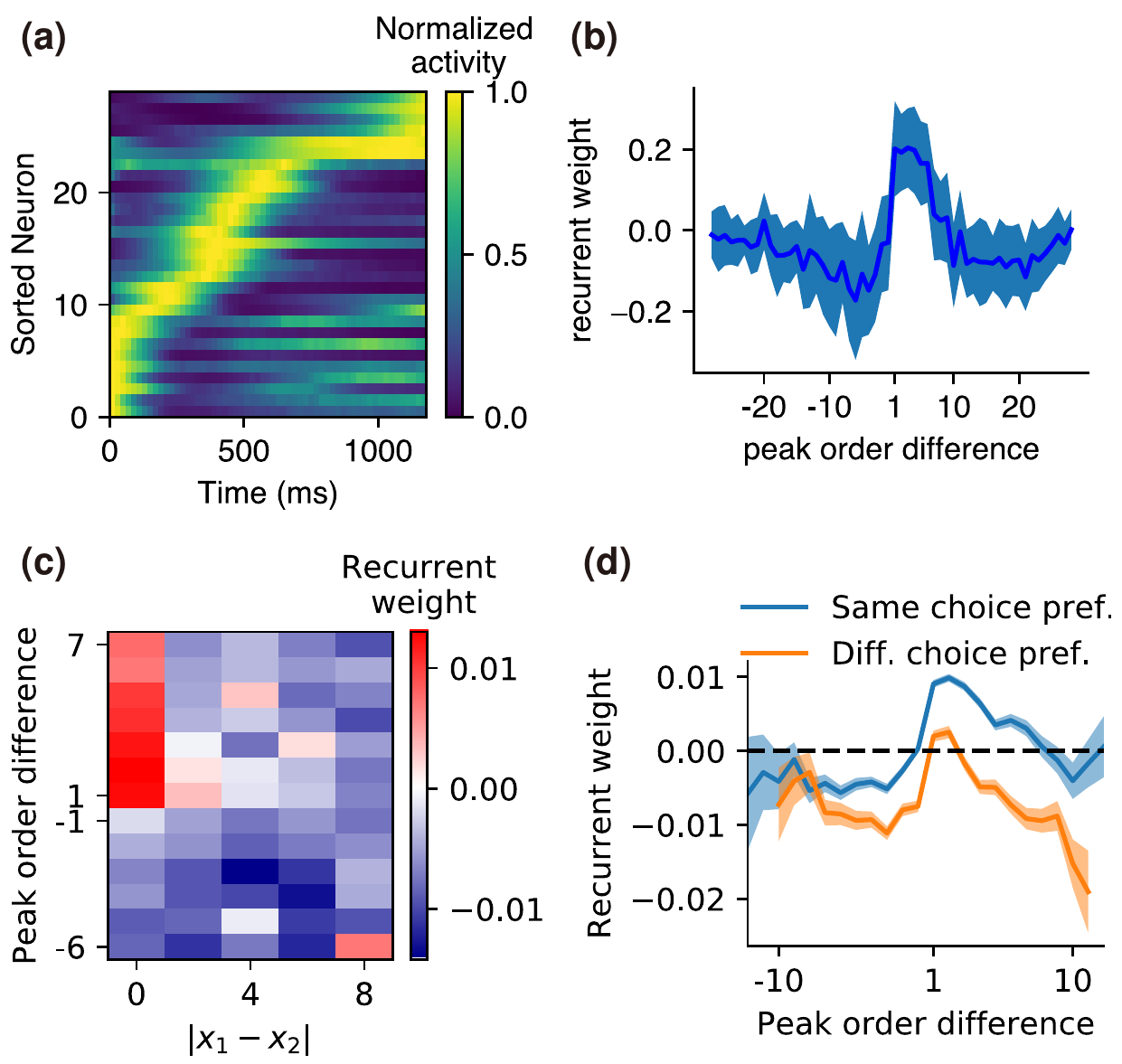}
	
	\protect\protect\protect\caption{\textbf{\label{fig:network_structure}Sequential activity and network
			structure. }(\textbf{a}) An example of neuronal activity (with maximum
		normalized to 1) in the perception epoch of IP task, sorted according
		to peak time. (\textbf{b}) Mean (solid line) and s.d. (shaded belt)
		of the recurrent weights as a function of the peak order difference
		between post- and pre-synaptic neurons in the perception epoch of
		IP. (\textbf{c}) Recurrent weight as a function of the difference
		$|x_{1}-x_{2}|$ between the preferred spatial locations of post-
		and pre-synaptic neurons and their peak order difference in the perception
		epoch of t-SR. (\textbf{d}) Recurrent weight as a function of peak
		order difference in the sequence of neurons with the same (blue)
		or different (orange) preferred decision choices in the perception
		epoch of t-DM. Shaded belt indicates s.e.m. See the sequential activity
		and network structure in other epochs of the four timing tasks of
		\textbf{Fig. \ref{fig:Model-setup}b,c} in \textbf{Fig. S9}.}
\end{figure}

A common feature of the network dynamics in all the epochs of the
four timing tasks above was neuronal sequential firing (\textbf{Fig.
\ref{fig:network_structure}a} and \textbf{Fig. S9a-c}). We ordered the peak
firing time of the neurons, and then measured the recurrent weight
as a function of the order difference between two neurons. We found,
on average, stronger connections from earlier- to later-peaking neurons
than from later- to earlier-peaking neurons (\textbf{Fig. \ref{fig:network_structure}b} and \textbf{Fig. 
S9d-f}) \cite{Rajan_2016,Hardy_2018a,Orhan_2019}. To study the network
structure that supported the coding orthogonality of temporal flow
and non-temporal information in the perception and production epochs
of t-SR (or t-DM), we classified the neurons into groups according
to their preferred spatial location (or decision choice). Given a
neuron $i$ and a group $\mathcal{G}$ of neurons ($i$ may or may
not belong to $\mathcal{G}$), we ordered their peak times, and investigated
the recurrent weight from $i$ to each neuron of $\mathcal{G}$ (except
$i$ itself if $i\in\mathcal{G}$), see Methods for details. In this
way, we studied the recurrent weight $w(o_{post}-o_{pre},|x_{post}-x_{pre}|)$
as a function of the difference $o_{post}-o_{pre}$ between the peak
orders of post- and pre-synaptic neurons and the difference $|x_{post}-x_{pre}|$
of their preferred non-temporal information (\textbf{Fig. \ref{fig:network_structure}c,
d}). In t-SR, firstly, $w(o_{post}-o_{pre},0)$ exhibited similar
asymmetry as that in IP (\textbf{Fig. \ref{fig:network_structure}b}),
positive if $o_{post}-o_{pre}>0$ and negative if $o_{post}-o_{pre}<0$,
which drove sequential activity. Secondly, $w(1,|x_{post}-x_{pre}|)$
decreased with $|x_{post}-x_{pre}|$, and became negative when $|x_{post}-x_{pre}|$
was large enough (\textbf{Fig. \ref{fig:network_structure}c}). Together,
the network of t-SR can be regarded as of several feedforward sequences,
with two sequences exciting or inhibiting each other depending on
whether their spatial preferences are similar or far different. The
sequential activity coded the flow of time, and the short-range excitation
and long-range inhibition maintained the spatial information \cite{Lim_2014}.
Similar scenario also existed in the network of t-DM (\textbf{Fig.
\ref{fig:network_structure}d}), where the sequential activity coded
the flow of time, and the inhibition between the sequences of different
decision preferences provided the mutual inhibition necessary for
making decisions \cite{Wang_2002}.

The scenario that feedforward structure hidden in recurrent connections
drives sequential firing has been observed in a number of modeling
works \cite{Rajan_2016,Hardy_2018a,Orhan_2019}. Our work extends
this scenario to the interaction of multiple feedforward sequences,
which can code temporal flow and non-temporal information simultaneously. 

\subsection*{Understanding the strong temporal signals in non-timing tasks}

We have shown that in the perception and production epochs of t-SR
and t-DM, when the network is required to record the temporal flow
and maintain the non-temporal information simultaneously, neuronal
temporal profiles exhibit similarity across non-temporal information
(\textbf{Fig. \ref{fig:t-SR}b} and \textbf{Fig. S5a}) and the subspaces
coding temporal flow and non-temporal information are orthogonal
with small mixed variance (\textbf{Fig. \ref{fig:t-SR}d, e, i,
j}). Interestingly, in tasks which do not require temporal information
to perform, such profile similarity, orthogonality and small mixed
variance were also experimentally observed \cite{Machens_2010,Kobak_2016}.
Moreover, the time-dependent variance explained more than 75\% of
the total variance in some non-timing tasks \cite{Machens_2010,Kobak_2016}.
It would be interesting to ask why non-timing tasks developed so
strong temporal signals, thereby understanding the factors that facilitate
the formation of time sense of animals.

\begin{figure*}
	\includegraphics[scale=0.65]{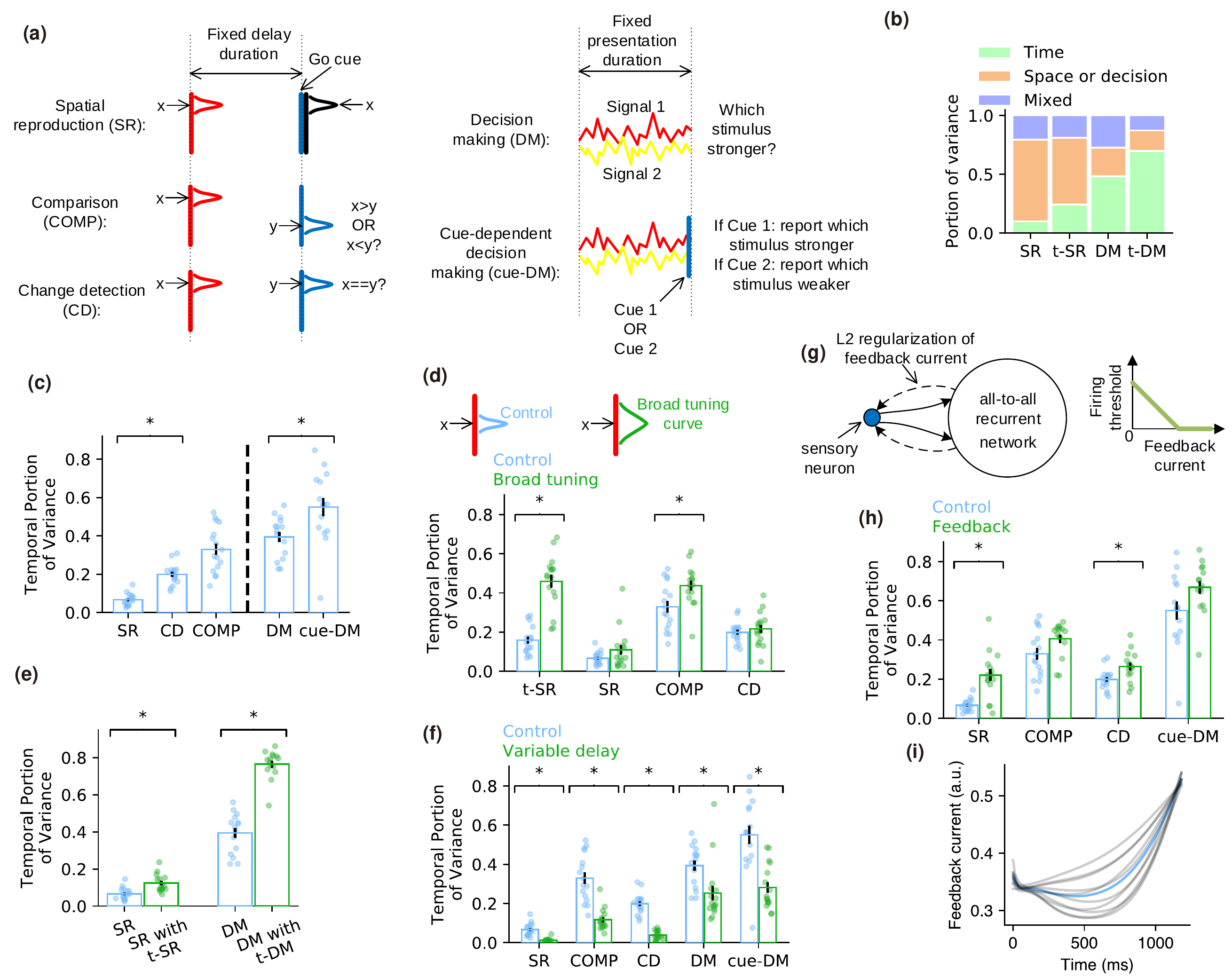}
	
	\protect\protect\caption{\label{fig:non-timing_task}\textbf{Understanding the strong temporal
			signals in non-timing tasks. }(\textbf{a}) Schematic of the non-timing
		tasks we studied. (\textbf{b}) Bar charts show how the total signal
		variance is split among temporal information, non-temporal information
		and the residual variance unexplained by temporal and non-temporal
		information in SR, t-SR, DM and t-DM. (\textbf{c}) The portion of
		total variance explained by temporal signals in non-timing tasks.
		Error bars represent mean $\pm$ s.e.m. across training configurations.
		Each dot corresponds to the value in a training configuration. Asterisk
		indicates significant difference at $p<0.05$ (two-sided Welch's
		t test). (\textbf{d}) The portion of time-dependent variance before
		(blue) and after (green) broadening the tuning curves of the sensory
		neurons. (\textbf{e}) The portion of time-dependent variance in SR
		or DM, when the network is trained on SR or DM only (blue), or trained
		on t-SR or t-DM concurrently (green). (\textbf{f}) The portion of
		time-dependent variance in fixed-delay (blue) or variable-delay (gree)
		tasks. (\textbf{g}) Left: schematic of the feedback connections (dashed
		arrows) to the sensory neurons (blue dot) to study anticipatory attention.
		Solid arrows represent feedforward connections. Right: firing threshold
		of sensory neuron decreases with feedback current until zero. (\textbf{h})
		The portion of time-dependent variance before (blue) and after (green)
		adding feedback to sensory neurons. (\textbf{i}) Feedback current
		as a function of time in the delay epoch of SR. Blue line: mean value.
		Gray lines: individual training configurations.}
\end{figure*}

First of all, before we studied the reasons for the strong temporal
signals observed in non-timing tasks, we studied how the requirement
of temporal processing influences the temporal signal strength. To
this end, we studied spatial reproduction task (SR), where the network
was to reproduce the spatial location immediately after a fixed delay
(\textbf{Fig. \ref{fig:non-timing_task}a, left column, first row}),
and decision making task (DM), where the network was to decide which
stimulus was stronger immediately after the presentation of two stimuli
(\textbf{Fig. \ref{fig:non-timing_task}a, right column, first row}).
Unlike t-SR (or t-DM) (\textbf{Fig. \ref{fig:Model-setup}c}), SR
(or DM) did not require the network to record time between the two
pulses (or during the presentation of the two stimuli). We used $p_{t}=\text{Var}_{i,t}(\langle r_{i}(t,x)\rangle_{x})/\text{Var}_{i,t,x}(r_{i}(t,x))$
to be the portion of time-dependent variance in the total variance
$\text{Var}_{i,t,x}(r_{i}(t,x))$, with $r_{i}(t,x)$ being the firing
rate of the $i$th neuron at time $t$ and non-temporal information
$x$. We compared the portions $p_{t}(\text{t-SR})$ and $p_{t}(\text{t-DM})$
in the perception epochs of t-SR and t-DM with the portion $p_{t}(\text{SR})$
in the delay epoch of SR and that $p_{t}(\text{DM})$ during the
presentation of stimuli in DM. We found that $p_{t}(\text{SR})<p_{t}(\text{t-SR})$
and $p_{t}(\text{DM})<p_{t}(\text{t-DM})$ (\textbf{Fig. \ref{fig:non-timing_task}b}).
Therefore, temporal signals are stronger in timing tasks. However,
even in the two timing tasks t-SR and t-DM we studied, the portion
of time-dependent variance was smaller than the portion (75\%) experimentally
observed in non-timing tasks \cite{Machens_2010,Kobak_2016} (\textbf{Fig.
\ref{fig:non-timing_task}b}). Therefore, there should exist other
factors than the timing requirement that are important to the formation
of temporal signals.

Specifically, we studied the following four factors: (1) temporal
complexity of task, (2) overlap of sensory input, (3) multi-tasking,
and (4) timing anticipation.

\textit{Temporal complexity of task}. Temporal complexity measures
the complexity of spatio-temporal patterns that the network receives
or outputs in a task \cite{Orhan_2019}. To test the influence of
temporal complexity on the strength of temporal signals, we designed
comparison (COMP) and change detection (CD) tasks that enhanced the
temporal complexity of SR and cue-dependent decision making (cue-DM)
task that enhanced the temporal complexity of DM (\textbf{Fig. \ref{fig:non-timing_task}a}).
In COMP, the network was to report whether the spatial coordinate
of the stimulus presented before the delay was smaller or larger
than that of the stimulus presented after the delay, in consistent
with its vibrotactile version \cite{Machens_2010} (\textbf{Fig.
\ref{fig:Model-setup}d}). In CD, the network was to report whether
the two stimuli presented before and after the delay were the same
\cite{Rossi-Poola_2019}. In cue-DM, the network was to report the
index of the stronger or weaker stimulus, depending on the cue flashed
at the end of the presented stimuli. COMP and CD have higher temporal
complexity than SR because the output not only depends on the stimulus
before the delay but also the stimulus after the delay. Similarly,
cue-DM has higher temporal complexity than DM because the output
also depends on the cue. We found that $p_{t}(\text{SR})<p_{t}(\text{CD})$,
$p_{t}(\text{SR})<p_{t}(\text{COMP})$ and $p_{t}(\text{DM})<p_{t}(\text{cue-DM})$,
suggesting that temporal complexity increases the portion of time-dependent
variance (\textbf{Fig. \ref{fig:non-timing_task}c}). It has been
empirically found that the task temporal complexity increases the
temporal fluctuations in neuronal sequential firing \cite{Orhan_2019}.
Here we showed that the temporal fluctuation of the average neuronal
activity $\{\langle r_{i}(t,x)\rangle_{x}\}_{i}$ over non-timing
information also increases with the temporal complexity of the task.
The result $p_{t}(\text{SR})<p_{t}(\text{COMP})$ is consistent with
the experimental observation that the population state varied more
with time in COMP than in SR \cite{Murray_2017}.

\textit{Overlap of sensory input}. Suppose the population states
of the sensory neurons in response to two stimuli $x_{1}$ and $x_{2}$
are respectively $\mathbf{s}_{1}$ and $\mathbf{s}_{2}$. If $\mathbf{s}_{1}$
and $\mathbf{s}_{2}$ have high overlap, then the evolution trajectories
of the recurrent network in response to $x_{1}$ and $x_{2}$ should
be close to each other. In this case, the variance of the trajectories
induced by the stimulus difference is small, and the time-dependent
variance explains a large portion of the total variance. To test
this idea, we broadened the Gaussian tuning curves of the sensory
neurons in t-SR, SR, COMP and CD tasks, and found increased portion
of time-dependent variance (\textbf{Fig. \ref{fig:non-timing_task}d}).

\textit{Multi-tasking}. The brain has been well trained on various
timing tasks in everyday life, so the animal may also have a sense
of time when performing non-timing tasks, which increases the time-dependent
variance. To test this hypothesis, we trained networks on t-SR and
SR concurrently, so that the network could perform either t-SR or
SR indicated by an input signal. \cite{Yang_2019}. We also trained
t-DM and DM concurrently. We only considered these two task pairs
because the two tasks in each pair share the same number and type
of inputs and outputs (except for a scalar Go-cue input in t-SR and
t-DM), hence they do not require any changes in the network architecture.
We found that both $p_{t}(\text{SR})$ and $p_{t}(\text{DM})$ were
larger in networks that were also trained on timing tasks than in
networks trained solely on non-timing tasks (\textbf{Fig.\ref{fig:non-timing_task}e}).

\textit{Timing anticipation.} In the working memory experiments that
observed strong time-dependent variance \cite{Machens_2010,Kobak_2016},
the delay period had fixed duration. This enabled the animals to learn this duration after long-term training and
predict the end of the delay, thereby getting
ready to take actions or receive new stimuli toward the end of the
delay. If the delay period is variable, then the end of the delay will no longer be predictable. We found that the temporal signals in fixed-delay tasks were stronger than those in
variable-delay tasks (\textbf{Fig. \ref{fig:non-timing_task}f}),
which suggests that timing anticipation is a reason for strong temporal
signals. A possible functional role of timing anticipation is anticipatory
attention: a monkey might pay more attention to its finger or a visual
location when a vibrotactile or visual cue was about to come toward
the end of the delay to increase its sensitivity to the stimulus.
To study the influence of this anticipatory attention to the formation of temporal signals, we supposed feedback
connections from the recurrent network to the sensory neurons in
our model (\textbf{Fig. \ref{fig:non-timing_task}g}). Feedback currents
could reduce the firing thresholds of sensory neurons through disinhibition
mechanism \cite{Hattori_2017}. We also added L2 regularization on
the feedback current (\textbf{Fig. \ref{fig:non-timing_task}g})
to reduce the energy cost of the brain (see Methods). After training,
the feedback current stayed at a low level to reduce the energy cost,
but became high when the cue was about to come to increase the sensitivity
of the network (\textbf{Fig. \ref{fig:non-timing_task}i}). We found
that adding this feedback mechanism increased the portion of time-dependent
variance (\textbf{Fig. \ref{fig:non-timing_task}h}), because the
feedback current, which increased with time toward the end of the
delay (\textbf{Fig. \ref{fig:non-timing_task}i}), provided a time-dependent
component of the population activity.

Collectively, other than the timing requirement in timing tasks,
we identified four possible factors that facilitate the formation
of strong temporal signals: (1) high temporal complexity of tasks;
(2) large sensory overlap under different stimuli; (3) transfer of
timing sense due to multi-tasking; (4) timing anticipation.

\section*{Discussion}

In summary, neural networks perceive time intervals through stereotypical
dynamic trajectories, maintain time intervals by attractor dynamics
in a complementary monotonic coding scheme, and perform interval
production or comparison by scaling evolution speed. Temporal and
non-temporal information are coded in orthogonal subspaces with small
mixed variance, which facilitates decoding generalization. The network
structure after training exhibits multiple feedforward sequences
that mutually excite or inhibit depending on whether their preferences
of non-temporal information are similar or not. We identified four
possible factors that facilitate the formation of strong temporal
signals in non-timing tasks: temporal complexity of task, overlap
of sensory input, multi-tasking and timing anticipation.

\subsection*{Perception and production of time intervals}

In the perception epoch, the network evolved along a stereotypical
trajectory after the first pulse (\textbf{Fig. \ref{fig:IP_task}b,
c}). Consistently, some neurons in the prefrontal cortex and striatum
prefer to peak their activities around specific time points after
an event \cite{Jin_2009}. In the brain, such stereotypical trajectory
may not only formed by neuronal activity state, but also synaptic
state such as slow synaptic current or short-term plasticity. The
temporal information coded by the evolution of synaptic state can
be read out by the network activity in response to a stimulus \cite{Buonomano_2000,Karmarkar_2007,Perez_2018}. 

We also studied the trajectory speed with time during the perception epoch. We found that after a transient period, the speed in IP task stayed around a constant value (\textbf{Fig. S2a}), the speed in IC and t-DM tasks increased with time (\textbf{Figs. S3c, S5c}), and the speed in t-SR task decreased with time (\textbf{Fig. S4b}). Therefore, we did not make any general conclusion on the trajectory speed when the network perceiving time intervals.

The temporal scaling when producing or comparing intervals has been
observed in animal experiments \cite{Wang_2018,Remington_2018,Mendoza_2018}.
A possible reason why temporal scaling exhibit in both the production
epoch of IP and the stimulus2 epoch of IC (\textbf{Fig. \ref{fig:IP_task}l-o} and 
\textbf{Fig. S3i-n}) is that both epochs require the network to compare the currently
elapsed time with the time interval maintained in working memory.
In IC, the decision choice is switched as soon as the trajectory
has passed the critical point at which the elapsed time $t$ equals
to the maintained interval $T$; in IP, the network is required to
output a movement as soon as $t=T$: both tasks share a decision-making process around the $t=T$ time point. This temporal scaling enables
generalizable decoding of the portion $t/T$ of the elapsed time
(\textbf{Figs. S7g, 8g}), which enables people to identify the same
speech or music played at different speeds \cite{Gutig_2009,Goudar_2018}. 

When the to-be-produced interval $T$ gets changed, the trajectory
in the perception epoch is truncated or lengthened (\textbf{Fig.
\ref{fig:IP_task}b-d}), whereas the trajectory in the production
epoch is temporally scaled (\textbf{Fig. \ref{fig:IP_task}l-o}).
This difference helps us to infer the psychological activity of the
animal. For example, in the fixed-delay working memory task, when
the delay period was changed, the neuronal activity during the delay
was temporally scaled \cite{Machens_2010}. This implies that the
animals had already learned the duration of the delay, and were actively
using this knowledge to anticipate the coming stimulus, instead of
passively perceiving time. However, this anticipation was not feasible
before the animal had learned the delay duration. Therefore, we predict
that at the beginning of training, the animal perceived time using
stereotypical trajectory, and the scaling phenomenon gradually emerged
during training.

\subsection*{Combination of temporal and non-temporal information}

Temporal and non-temporal information are coded orthogonally with
small mixed variance (\textbf{Fig. \ref{fig:t-SR}}). Physically,
time is consistently flowing, regardless of the non-temporal information;
and much information is also invariant with time. The decoding generalizability
resulted from this coding geometry (\textbf{Fig. \ref{fig:Decoding-generalizability}})
helps the brain to develop a shared representation of time across
non-temporal information or a shared representation of non-temporal
information across time using a fixed set of readout weights. Decoding
generalizability of non-temporal information across time has been
studied in working memory tasks \cite{Murray_2017,Stokes_2013},
and has been considered as an advantage of working memory models
in which information is maintained in stable, or a stable subspace
of, neuronal activity \cite{Murray_2017,Druckmann_2012}. Here we
showed that with this geometry, such advantage also exists for reading
out temporal information.

Interestingly, the orthogonality and small mixed variance have been
experimentally observed in non-timing tasks \cite{Machens_2010,Kobak_2016},
so their formation seems not to depend on the timing task requirement.
Consistently, in the delay epoch of t-SR and t-DM, although the network
needed not to record the temporal flow to perform the tasks, temporal
flow and non-temporal information were still coded orthogonally with
small mixed variance (\textbf{Fig. S10a-d}). By comparing the orthogonality
and mixed variance in the perception epoch of t-SR and t-DM with
those in the non-timing tasks (\textbf{Fig. \ref{fig:non-timing_task}a}),
we found that timing task requirement did not influence the orthogonality,
but generally reduced the mixed variance (\textbf{Fig. S10e, f}).
The network structure in non-timing tasks also exhibited interacting
feedforward sequences (\textbf{Fig. S10g-k}).

\subsection*{Strong temporal signals in non-timing tasks}

Our results concerning the various factors that affect the strength
of temporal signals in non-timing tasks lead to testable experimental
predictions. The result of sensory overlap (\textbf{Fig. \ref{fig:non-timing_task}d})
implies that sensory neurons with large receptive fields are essential
to the strong temporal signals. The result of multi-tasking (\textbf{Fig.
\ref{fig:non-timing_task}e}) implies that animals better trained
on timing or music have stronger temporal signals when performing
non-timing tasks. The result of temporal complexity (\textbf{Fig.
\ref{fig:non-timing_task}c}) implies that animals have stronger
temporal signals when performing tasks with higher temporal complexity,
which is consistent with some experimental clues \cite{Murray_2017}.
The result of timing anticipation (\textbf{Fig. \ref{fig:non-timing_task}f}),
consistently with Ref. \cite{Orhan_2019}, implies that if the appearance
of an event is unpredictable, then the temporal signals should be
weakened. Besides anticipatory attention (\textbf{Fig. \ref{fig:non-timing_task}g}),
anticipation may influence the temporal signals through other mechanisms.
In the fixed-delay comparison task (\textbf{Fig. \ref{fig:Model-setup}d}),
suppose a stimulus $a$ appeared before the delay, then both the
population firing rate and the information about $a$ in the population
state increase toward the end of the delay period \cite{Barak_2010}.
It is believed that this is because the information about $a$ was
stored in short-termly potentiated synapses in the middle of the
delay to save the energetic cost of neuronal activity, while got
retrieved into the population state near the end of the delay period
to facilitate information manipulation \cite{Masse_2019}. This storing
and retrieving process may also be a source of temporal signals.

\subsection*{Interval and beat based timing}

We have discussed the processing of single time intervals using our
model. However, recent evidences imply that the brain may use different
neural substrates and mechanisms to process regular beats from single
time intervals \cite{Teki_2011,Gamez_2019}. Dynamically, in the medial
premotor cortices, different regular tapping tempos are coded by
different radii of circular trajectories that travel at a constant
speed \cite{Gamez_2019}, which is different from the stereotypical
trajectory or speed scaling scenario revealed in our model (\textbf{Fig.
\ref{fig:IP_task}b-d, l-o}). The distribution of the preferred intervals
of the tapping-interval-tuned neurons is wide, peaking around 850
ms \cite{Merchant_2013_b,Bartolo_2014}, which is also different from
the complementary monotonic tuning scenario in our model (\textbf{Fig.
\ref{fig:IP_task}j, k}). Additionally, humans tend to use a counting
scheme to estimate single time intervals when the interval duration
is longer than 1200 ms \cite{Grondin_1999}, which implies that the
beat-based scheme is mentally used to reduce the estimation error
of single long intervals even without external regular beats. However,
after we trained our network model to produce intervals up to 2400
ms, it processed intervals between 1200 ms and 2400 ms in similar
schemes (\textbf{Fig.
	S11}) to that illustrated in \textbf{Fig. \ref{fig:IP_task}}  for intervals below 1200 ms. All these results suggest the limitation of our model to explain
beat-based timing. Modeling work on beat-based timing is the task
of future research. 

\section*{Methods}

Methods and Figs. S1 to S11 are provided in supplementary information.
In the method section, we present the details of our computational
model, including the network structure, the tasks to be performed
and the methods we used to train the network model. We also present
the details to analyze the interval-coding scheme in the delay and
production epochs (\textbf{Fig. \ref{fig:IP_task}}), the coding
combination of temporal and non-temporal information (\textbf{Figs.
\ref{fig:t-SR}, \ref{fig:non-timing_task}b}), decoding generalization
(\textbf{Fig. \ref{fig:Decoding-generalizability}}) as well as the
firing and structural sequences (\textbf{Fig. \ref{fig:network_structure}}).
We also explain the relationship between the monotonic coding at
the end of the delay epoch and the low dimensionality of the attractor
(\textbf{Fig. \ref{fig:IP_task}g, k}), as well as the geometric
meaning of mixed variance (\textbf{Fig. \ref{fig:t-SR}}) in supplementary
information. Computer code is available from \url{https://github.com/zedongbi/IntervalTiming}.

\section*{Acknowledgment}

This work was supported by Hong Kong Baptist University (HKBU) Strategic Development Fund and HKBU Research Committee and Interdisciplinary Research Clusters Matching Scheme 2018/19 (RC-IRCMs/18-19/SCI01). This research was conducted using the resources of the High-Performance Computing Cluster Centre at HKBU, which receives funding from the RGC and HKBU.

\bibliographystyle{ieeetr}
\bibliography{reference}

\end{document}



\maketitle

\section{Method}

\subsection{Network details}

We adopted a discrete-time formulation of network dynamics, in which
\begin{equation}
\mathbf{x}_{t}=\mathbf{W}^{rec}\mathbf{r}_{t-1}+\mathbf{W}^{in}\mathbf{u}_{t}+\mathbf{W}^{in,att}[\mathbf{u}_{t}^{att}-\theta^{att}]_{+}+\mathbf{b}+\sqrt{2\sigma_{rec}^{2}}\text{N}(0,1),\label{eq:1}
\end{equation}
where $\mathbf{x}_{t}$, $\mathbf{r}_{t}$ and $\mathbf{u}_{t}$
are respectively the synaptic current, firing rate and network input
at time step $t$, $\mathbf{b}$ is the background input, $\mathbf{W}^{rec}$
is the recurrent weight, $\mathbf{W}^{in}$ is the input weight,
and $\sigma_{rec}$ is the strength of recurrent noise. We supposed
$\mathbf{r}_{t}=f(\mathbf{x}_{t})$, with $f(\cdot)$ being the softplus
current-rate transfer function, i. e. 
\begin{equation}
f(x)=\log(1+\exp(x)).\label{eq:2}
\end{equation}
Input $\mathbf{u}_{t}$ is also noisy, 
\begin{equation}
\mathbf{u}_{t}=\mathbf{u}_{signal}+\sqrt{2\sigma_{in}^{2}}\text{N}(0,1),\label{eq:input_noise}
\end{equation}
with $\sigma_{in}$ being the strength of input noise. $\mathbf{W}^{in,att}$,
$\mathbf{u}_{t}^{att}$ and $\theta^{att}$ are the quantities related
to the input units modulated by top-down attention. They are only
valid when studying the effect of anticipatory attention in non-timing
tasks (\textbf{Fig. 6g-i}). The model does not have these quantities
in the other tasks. $\mathbf{W}^{in,att}$ is the weight from the
attention-modulated units to the recurrent network, $\mathbf{u}_{t}^{att}$
is the input current to the attention-modulated units, and $\theta^{att}$
is the firing threshold of these units. The firing threshold is 
\begin{equation}
\theta^{att}=[\theta_{0}^{att}-\mathbf{W}^{fb,att}\mathbf{r}_{t}]_{+},\label{eq:feedback_threshold}
\end{equation}
with $\mathbf{W}^{fb,att}$ being positive feedback weight, so that
$\theta^{att}$ decreases with feedback current until to zero, starting
from $\theta_{0}^{att}=1.5$. Eq. \ref{eq:feedback_threshold} models
the disinhibitory effect of feedback connections \cite{Hattori_2017}.
Similar to $\mathbf{u}_{t}$, $\mathbf{u}_{t}^{att}$ is also noisy,
with the noise strength $\sigma_{in}^{2}$ (eq. \ref{eq:input_noise}).

Some previous studies started with a continuous-time formulation,
and obtained the discrete-time version using Euler method (omitting
the attention-modulated units): 
\begin{equation}
\mathbf{x}_{t}=(1-\alpha)\mathbf{x}_{t-1}+\alpha(\mathbf{W}^{rec}\mathbf{r}_{t-1}+\mathbf{W}^{in}\mathbf{u}_{t}+\mathbf{b}+\sqrt{2\alpha^{-1}\sigma_{rec}^{2}}\text{N}(0,1)),\label{eq:3}
\end{equation}
with $\alpha=\Delta t/\tau$ being the ratio of time step length
$\Delta t$ and membrane time constant $\tau$. In our study, we
effectively set $\alpha=1$, similarly as the scheme used in Ref.
\cite{Wang_2018,Orhan_2019}. We also set $\Delta t=20\text{ ms}$.
The output of the network is supposed to be 
\begin{equation}
z=\mathbf{W}^{out}\mathbf{r}+\mathbf{b}^{out},\label{eq:output}
\end{equation}
with the dimension of $z$ depending on tasks.

We set $\sigma_{in}=0.01$, $\sigma_{rec}=0.05$ when training the
network. After training, when plotting the neuronal activities in
the perception epoch (\textbf{Fig. 2b-d}), we kept $\sigma_{in}=0.01$,
$\sigma_{rec}=0.05$ so that the neuronal temporal profiles under
different durations of perception epoch did not fully overlap. When
doing the other analysis, we turned off the noises by default.

\subsection{Task details}

\subsubsection{Timing tasks}

\begin{figure}[tbph]
\centering \includegraphics[scale=0.6]{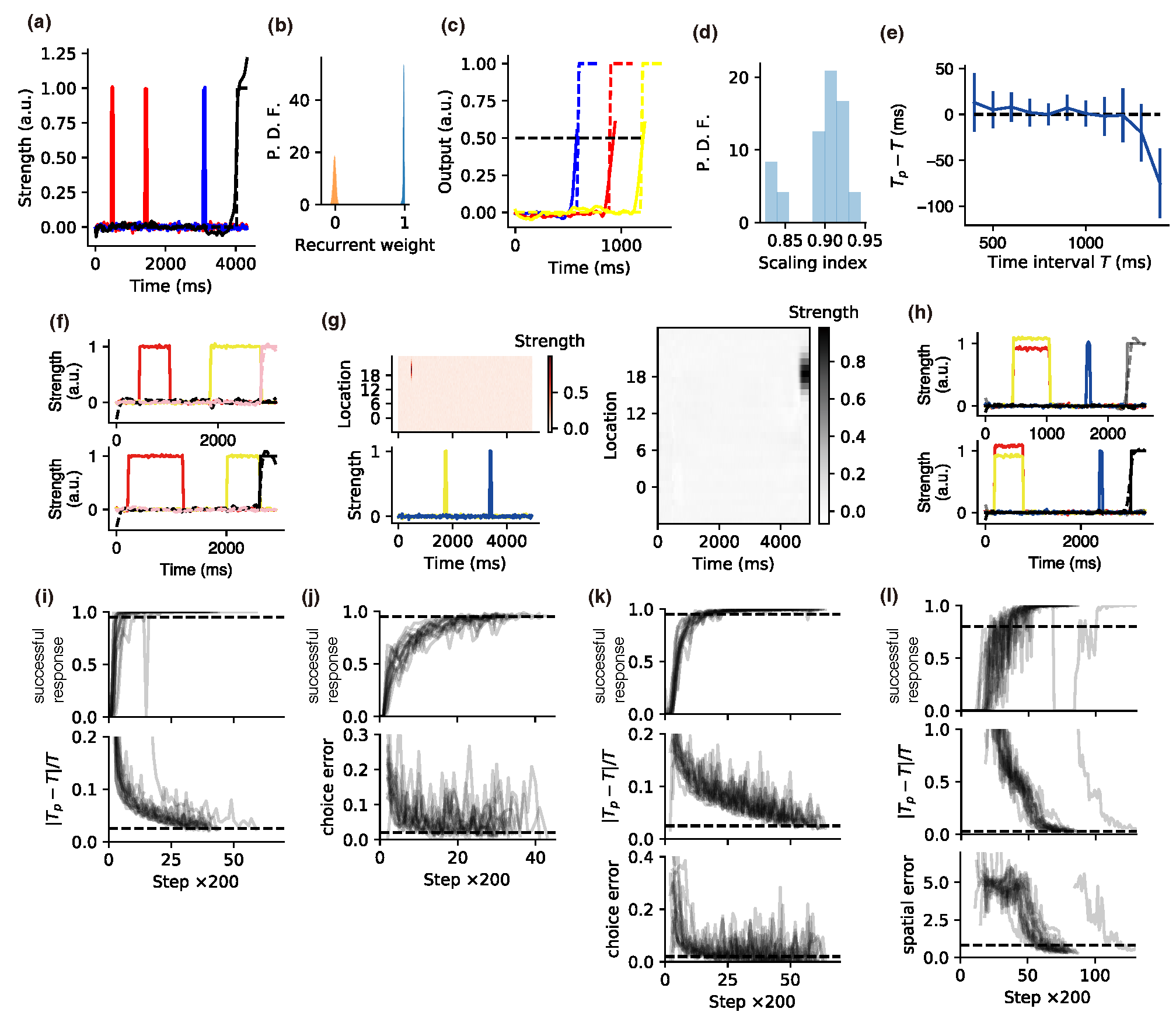}

\protect\protect\protect\caption{\textbf{\label{fig:Performance-after-training}Performance of the
network after training. }(\textbf{a-e}) Interval production (IP)
task. (\textbf{a}) An example of the input and output of the network
in IP. Red and blue lines: two input channels. Dashed black line:
target output. Solid black line: actual output. (\textbf{b}) Probability
distribution function (p.d.f) of self-connections (blue) and non-diagonal
connections (red) of the recurrent network after training. (\textbf{c})
Three examples of the output in the production epoch of IP, when
$T=600$ ms (blue), 900 ms (red) and 1200 ms (yellow). Dashed line:
target output. Solid line: actual output. The horizontal dashed black line indicates the threshold that the network is regarded to generate a movement in the production epoch when the output rises across this threshold.  (\textbf{d}) Distribution
of the scaling index of the output across training configurations
in the production epoch of IP. (\textbf{e}) The difference between
the produced time interval $T_{p}$ and the interval $T$ between
the first two pulses in IP as a function of $T$. Error bar means
standard deviation over 16 training configurations. During training,
we set $T\in[400\text{ ms},1400\text{ ms}]$. This panel shows that
if after training we set $T$ to be close to 400 ms, $T_{p}$ tends
to be larger than $T$; whereas if we set $T$ to be close to 1400
ms, $T_{p}$ tends to be smaller than $T$. Therefore, by default,
we set $T\in[600\text{ ms},1200\text{ ms}]$ for data analysis after
training to reduce the bias of $T_{p}$. (\textbf{f}) Two examples
of interval discrimination (IC) task. Upper: the case when the duration
of the first stimulus is shorter than that of the second stimulus.
Lower: the case when the duration of the first stimulus is longer
than that of the second stimulus. Red and yellow lines: two input
channels. Dashed black and pink lines: two channels of target output.
Solid black and pink lines: two channels of actual output. (\textbf{g})
An example of timed spatial reproduction (t-SR) task. Left upper:
the pulse with location information from the first input channel.
Left lower: the pulses from the second (yellow) and third (blue)
input channels. Right: actual output. (\textbf{h}) Two examples of
timed decision making (t-DM) task. Upper: when the input from the
first channel (red) is weaker than the input from the second channel
(yellow), i.e., $c<0$. Lower: when $c>0$. (\textbf{i-l}) Performance of the network during training. (\textbf{i}) Performance
of the network during the training of IP, quantified by the probability
to successfully produce time interval (upper) and the relative error of
the produced interval (lower). Gray lines indicate individual training
configurations. Training stopped as soon as both quantities reach
the criterion (horizontal dashed lines). (\textbf{j}) Performance
of the network during the training of IC, quantified by the probability
to successfully output a choice (upper) and the probability of choice
error (lower). (\textbf{k}) Performance of the network during the
training of t-SR, quantified by the probability to successfully produce time
interval (upper), the relative error of the produced interval (middle)
and the spatial error of the output. (\textbf{l}) Performance of
the network during the training of t-DM, quantified by the probability
to successfully produce time interval (upper), the relative error of the
produced interval (middle) and the probability of choice error (lower). }
\end{figure}

\textit{Interval production task (IP).} The network received from
2 input units: from one came the two pulses that defined the time
interval, and from the other came the Go cue. The interval between
the beginning of the simulation and the onset of the first pulse
was 
\begin{equation}
T_{start}\sim U(60\text{ ms},500\text{ ms}),\label{eq:T_start}
\end{equation}
where $U(t_{1},t_{2})$ is a uniform distribution between $t_{1}$
and $t_{2}$. The interval between the offset of the first pulse
and the onset of the second pulse was 
\begin{equation}
T\sim U(400\text{ ms},1400\text{ ms}).\label{eq:T}
\end{equation}
Note that we set the range of $T$ to be $[400\text{ ms},1400\text{ ms}]$
during training, but after training, we only investigated the performance
of the network when $T\in[600\text{ ms},1200\text{ ms}]$. The reason
is that there were boundary effects if, after training, $T$ took
a value close to 400 ms or 1400 ms: if $T$ was close to 400 ms,
then the time interval produced by the network was biased to be larger
than $T$; whereas if $T$ was close to 1400 ms, then the produced
interval was biased to be smaller than $T$. Such biases were weak
if $T$ took a middle value (\textbf{Fig. \ref{fig:Performance-after-training}e}).

The interval between the offset of the second pulse and the onset
of the Go cue (i. e. , the delay period) was 
\begin{equation}
T_{delay}\sim U(600\text{ ms},1600\text{ ms}).\label{eq:T_delay}
\end{equation}
All input pulses (including the two pulses that defined the time
interval, and the Go cue) lasted for 60 ms, and had strength 1. Input
units stayed at 0 when there were no pulses.

The target output was a scalar. It stayed at zero from the beginning,
jumped to 1 at time $T$ after the offset of the Go cue, and kept
at 1 until the end of the simulation at 300ms afterwards.

\textit{Interval comparision task (IC).} The network received two
successive long-lasting stimuli respectively from two input units.
The first stimuli, which came from the first unit, started at time
$T_{start}$ after the beginning of the simulation, and lasted for
duration $T_{1}$. Then after a delay interval $T_{delay}$ , the
second stimuli, which came from the second unit, started, and lasted
for duration $T_{2}$.

\begin{equation}
T_{start}\sim U(60\text{ ms},500\text{ ms}),\quad T_{1}\sim U(400\text{ ms},1400\text{ ms}),\quad T_{delay}\sim U(600\text{ ms},1600\text{ ms}),\quad T_{2}\sim U(400\text{ ms},1400\text{ ms})\label{eq:T_discrim}
\end{equation}
All the input stimuli had strength 1. Input units stayed at 0 when
there were no stimuli.

The target outputs were two scalars $\hat{z}_{0}$ and $\hat{z}_{1}$.
Both stayed at zero from the beginning. If $T_{1}>T_{2}$, then $\hat{z}_{0}$
jumped to 1 at the offset of the second stimulus, and stayed at 1
until the end of the simulation at 300 ms afterwards. Otherwise,
$\hat{z}_{1}$ jumped to 1 at the offset of the second stimulus.

\textit{Timed spatial reproduction task (t-SR).}\textbf{ }The network
successively received three pulses from three input channels. The
first channel was a line that coded spatial locations. This line
contained 32 units, whose preferred directions were uniformly spaced
from -6 to 25. For unit $i$ with preferred location $y_{i}$, its
activity in a pulse with location $x$ was 
\begin{equation}
A_{in}(t)\exp[-\frac{1}{2}(\frac{|y_{i}-x|}{2})^{2}],\label{eq:spatial_strength}
\end{equation}
where $A_{in}(t)=1$ during the presentation of the pulse and $A_{in}(t)=0$
at the other time. In our simulation, the spatial locations of the
stimuli were uniformly drawn from 0 to 19. The second and third channels
were both scalar inputs. The pulse from the second channel defined
the time interval to be remembered together with the pulse from the
first channel. The pulse from the third channel acted as Go cue.
$T_{start}$, $T$ and $T_{delay}$ were distributed similarly as
in IP (eqs. \ref{eq:T_start}-\ref{eq:T_delay}).

The target output was a line with 32 units, which represented response
location using similar tuning curves as the ones used for the input
line (eq. \ref{eq:spatial_strength}): 
\begin{equation}
\hat{z}_{i}=A_{out}(t)\exp[-\frac{1}{2}(\frac{|y_{i}-x|}{2})^{2}],\label{eq:target_output_ring-1}
\end{equation}
where the amplitude $A_{out}(t)$ stayed at zero from the beginning,
jumped to 1 at time $T$ after the offset of the Go cue, and stayed
at 1 until the end of the simulation at 300 ms afterwards.

\textit{Timed decision making task (t-DM).}\textbf{ }The network
received from three channels of scalar inputs. From the first two
channels came the stimuli whose strengths were to be compared with
each other, and from the last channel came the Go cue pulse. Starting
from the beginning of simulation, the first two channels were set
to 0 for duration $T_{start}$, and then jumped to $A_{1}$ and $A_{2}$
respectively; after $T$ time, these two channels were set to 0 again.
The Go cue pulse came at time $T_{delay}$ after the offset of the
first two channels. Here, 
\begin{equation}
A_{1}=\gamma+c,\quad A_{2}=\gamma-c,\label{eq:StimStrengthDM}
\end{equation}
where $\gamma$ was the average strength of these two stimuli and
was distributed as $\gamma\sim U(0.8,1.2)$, and $c$ measured the
strength difference of these two stimuli, and was distributed as
\begin{equation}
c\sim U(\{-0.08,-0.04,-0.02,-0.01,0.01,0.02,0.04,0.08\}),\label{eq:StimDiffDM}
\end{equation}
where $U(\{a_{1},a_{2},\cdots,a_{n}\})$ denotes a discrete uniform
distribution over the set $\{a_{1},a_{2},\cdots,a_{n}\}$. $T_{start}$,
$T$ and $T_{delay}$ were distributed similarly as in interval production
task (eqs. \ref{eq:T_start}-\ref{eq:T_delay}).

The target outputs were two scalars $\hat{z}_{0}$ and $\hat{z}_{1}$.
Both stayed at zero from the beginning. If $c>0$, then $\hat{z}_{0}$
jumped to 1 at time $T$ after the offset of the Go cue, and stayed
at 1 until the end of the simulation at 300ms afterwards. Otherwise,
$\hat{z}_{1}$ jumped to 1 at time $T$ after the offset of the Go
cue.

\subsubsection{Non-timing tasks: default settings}

\textit{Spatial reproduction task (SR).} The network received pulses
from two input channels. The first channel was a line that contained
32 units, coding spatial locations in the range $[-6,25]$ in the
way indicated by eq. \ref{eq:spatial_strength}. In our simulation,
the spatial locations of the stimuli were uniformly drawn from 0
to 19. The second channel is a scalar input. The duration $T_{delay}$
of the delay epoch between the first and second pulses was 1200 ms.
The target output was a line of 32 units (eq. \ref{eq:target_output_ring-1}),
which was to indicate the location of the first pulse immediately
after the second pulse.

\textit{Comparison task (COMP).} The network received pulses from
two input channels, both of which were lines that contained 32 units
successively gave two pulses to the network. The target outputs were
two scalars $\hat{z}_{0}$ and $\hat{z}_{1}$, which were to indicate
whether or not the spatial coordinate of the first pulse was larger
than that of the second pulse.

\textit{Change detection task (CD).} The network had the same structure
as that in COMP. Two scalar outputs were to indicate whether or not
the distance between the spatial locations of the two input pulses
was within 1.

\textit{Decision making task (DM).} The network received from two
channels of stimuli lasting for $T=1200$ ms. The two scalar outputs
were to indicate which stimulus was stronger immediately after the
ending of the two stimuli.

\textit{Cue-dependent decision making task (cue-DM).} The network
received from two channels of stimuli lasting for $T=1200$ ms. At
the 1140 ms after the presentation of the two stimuli, a two dimensional
one-hot vector lasting for 60ms was input from a third channel. Two
scalar outputs were to indicate the index of the stronger stimulus
or the index of the weaker stimulus according to the third channel.

\subsubsection{Non-timing tasks: studying the factors that influence the strength
of temporal signal}

To study the effect of the overlap of sensory input to the strength
of temporal signal in the delay epoch of SR, COMP and CD (\textbf{Fig.
6d}), we expanded the unit number in the line channels to 44 (default
is 32), and broadened the standard deviation of the tuning curves
(eq. \ref{eq:target_output_ring-1}) to 4 (default is 2). These units
coded spatial locations in the range -12 to 31. In our simulation,
the spatial locations of input stimuli were uniformly drawn from
0 to 19.

To study the effect of multi-tasking (\textbf{Fig. 6e}), we trained
the network on t-SR and SR concurrently, or on t-DM and DM concurrently.
The two tasks in each pair share the same input and output channels.
We used a one-hot vector from another two-dimensional input channel
to indicate which task should be performed \cite{Yang_2019}. The
network was to be able to perform either of the indicated task.

To study the effect of timing anticipation (\textbf{Fig. 6f}), we
trained the network to perform SR, COMP, CD, DM and cue-DM, with
the duration $T$ of the delay epoch (for SR, COMP and CD) or the
stimuli-presentation epoch (for DM and cue-DM) was randomly between
$[800\text{ ms},1600\text{ ms}]$. After training, we analyzed the
simulation results when $T=1200$ ms, and compared the results with
the cases after training the network with $T$ fixed at 1200 ms.
To study the the effect of anticipatory attention (\textbf{Fig. 6g-i}),
feedback was imposed on the second input channel of SR, COMP and
CD, and was imposed on the third channel of cue-DM. This means that
these input channels were modeled using the third term at the right-hand
side in eq. \ref{eq:1}, instead of the second term.

\subsection{Training details}

Training was performed to minimize a cost function using back-propagation
through time. Cost function was defined as 
\begin{equation}
C=\sum_{i}m_{i}(z_{i}-\hat{z}_{i})^{2},\label{eq:cost_fcn}
\end{equation}
where $i$ is the index of output units, $z_{i}$ is the actual output
defined by eq. \ref{eq:output}, $\hat{z}_{i}$ is the target output,
and $m_{i}$ is the mask. In all tasks, $m_{i}=0$ before the onset
of the first stimulus, and $m_{i}=1$ afterwards; therefore, only
the output after the onset of the first stimulus was constrained.
When studying the effect of anticipatory attention in non-timing
tasks (\textbf{Fig. 6g-i}), we added L2 regularization to feedback
current $\mathbf{I}^{fb}=\mathbf{W}^{fb,att}\mathbf{r}_{t}$ (see
eq. \ref{eq:feedback_threshold}), so that eq.\ref{eq:cost_fcn}
becomes $C=\sum_{i}m_{i}(z_{i}-\hat{z}_{i})^{2}+\beta_{fb}\frac{1}{N_{i,t}}\sum_{i,t}(I_{i,t}^{fb})^{2}$,
with $\beta_{fb}=10^{-4}$. This cost function was minimized using
Adam optimizer at learning rate 0.0005, with batch size 64 in each
training step. We trained 16 configurations to perform IP and IC
tasks, and trained 30 configurations to perform t-SR and t-DM tasks. Different
configurations were initialized using different random seeds. 

Before training, recurrent self-connections ($W_{ii}^{rec}$ in eq.
\ref{eq:3}) were initialized to 1, and other recurrent connections
were initialized as independent Gaussian variables with mean 0 and
standard deviation $0.3/\sqrt{N_{rec}}$, with $N_{rec}=256$ being
the number of recurrent units. This initialization strategy was used
in Ref. \cite{Orhan_2019}. The identity self-connections prevent
vanishing gradient during training \cite{Le_2015}, and the non-zero
off-diagonal recurrent connections induce sequential activity in
the network after training \cite{Orhan_2019}, so that the dynamics
of the network becomes comparable to experimental observations \cite{Goel_2016,MacDonald_2011,Hahnloser_2002,Pastalkova_2008,Harvey_2012}.
Output connections were initialized as independent Gaussian variable
with mean 0 and standard deviation $1/\sqrt{N_{rec}}$. Input connections
from the line input were initialized as variables drawn uniformly
from $[-1/\sqrt{2\sigma_{tuning}},1/\sqrt{2\sigma_{tuning}}]$, with
$\sigma_{tuning}$ being the standard deviation of the Gaussian tuning
curve (eq. \ref{eq:spatial_strength}), which was 2 by default and
4 when studying the effect of input overlap in non-timing tasks.
The input connections from the other channels were initialized as
variables drawn uniformly from $[-1/\sqrt{D_{channel}},1/\sqrt{D_{channel}}]$,
with $D_{channel}$ being the dimension of the input channel.

Every 200 training steps, we evaluated the performance of the network
using a batch of size 512, and stopped training as soon as the performance
of the network reached criterion (\textbf{Fig. \ref{fig:Performance-after-training}i-l}).
We introduced our criterion in t-SR and t-DM in details, the other
tasks shared similar criterion:

In t-SR, a time interval was considered to be produced if: (1) the
activities of all the 32 output units were below 0.2 before the offset
of the Go cue, (2) one of them went above 0.5 at some point $t_{p}$
before $T+300\text{ms}$ after the offset time $t_{off}^{cue}$ of
the Go cue. The produced interval was $T_{p}=t_{off}^{cue}-t_{p}$.
Output location at time $t_{p}$ was read out using a population
vector method (see the computer code in Ref. \cite{Yang_2019}).
Training was stopped as soon as (1) time intervals were produced
in over 95\% simulation trials, (2) the relative error of the produced
intervals $|T_{p}-T|/T<0.025$, (3) the output locations were on
average within 0.8 of the input locations.

In t-DM, a time interval was considered to be produced if: (1) the
activities of both output units $z_{0}$ and $z_{1}$ were below
0.2 before the offset of the Go cue, (2) one of them went above 0.5
at some time point $t_{p}$ before $T+300\text{ms}$ after the offset
$t_{off}^{cue}$ of the Go cue, whereas the other one stayed below
0.5. The produced interval was $T_{p}=t_{off}^{cue}-t_{p}$. In the
trials in which a time interval was produced, the decision was considered
to be correct if: when $c>0$ (or $c<0$), $z_{0}$ (or $z_{1}$)
went above 0.5 and $z_{1}$ (or $z_{0}$) kept below 0.5. Training
was stopped as soon as (1) time intervals were produced in over 96\%
of simulation trials, (2) the relative error of the produced intervals
$|T_{p}-T|/T<0.025$, (3) the decision error rate was smaller than
0.02.

\subsection{Data analysis}

\subsubsection{Types of neurons at the end of the delay epoch\label{sub:Types-of-neurons-delay-epoch}}

In IP or IC, we supposed $f_{i}(T)$ to be the activity of the $i$th
neuron at the end of the delay epoch as a function of the duration
$T$ of the perception (for IP) or stimulus1 (for IC) epoch. We picked
neurons that can be strongly activated at the end of the delay epoch,
namely the neurons whose $\max_{T\in[T_{min},T_{max}]}f_{i}(T)>\theta_{sa}$,
with $T_{min}=600\text{ ms}$ and $T_{max}=1200\text{ ms}$ respectively
being the minimal and maximal values of $T$ in our simulation, and
$\theta_{sa}=2$. Our results are not sensitive to the value of $\theta_{sa}$.
We classified $f_{i}(T)$ of the picked neurons into three types,
namely monotonically increasing (MoI), monotonically decreasing (MoD),
and non-monotonic (non-M) in the following way: We divided the range
of $T$ (i.e., $[T_{min},T_{max}]$) into four parts of the same
length, and calculated the mean value of $f_{i}(T)$ in these four
parts, say $f_{i}(\text{part 1})=\frac{4}{T_{max}-T_{min}}\int_{T_{min}}^{T_{min}+(T_{max}-T_{min})/4}f_{i}(T)\mathrm{d}T$,
$f_{i}(\text{part 2})=\frac{4}{T_{max}-T_{min}}\int_{T_{min+(T_{max}-T_{min})/4}}^{T_{min}+2(T_{max}-T_{min})/4}f_{i}(T)\mathrm{d}T$,
etc. If $f_{i}(\text{part 1})\le f_{i}(\text{part 2})\le f_{i}(\text{part 3})\le f_{i}(\text{part 4})$,
then neuron $i$ belongs to MoI type; if $f_{i}(\text{part 1})\ge f_{i}(\text{part 2})\ge f_{i}(\text{part 3})\ge f_{i}(\text{part 4})$,
then neuron $i$ belongs to MoD type; otherwise, neuron $i$ belongs
to non-M type. 

In t-SR, we supposed $g_{i}(T,x)$ to be the activity of the $i$th
neuron at the end of the delay epoch as a function of $T$ at a given
location $x$ of the first pulse. We picked neurons that can be strongly
activated at the end of the delay epoch (i.e., the neurons whose
$\max_{\{T,x\}}g_{i}(T,x)>\theta_{sa}$). We then defined $f_{i}(T)=\max_{x}g_{i}(T,x)$,
and classified neuron $i$ into MoI, MoD or non-M types according
to the monotonicity of $f_{i}(T)$ in the similar way to the IP or
IC case introduced above. Similarly, in t-DM, we classified neurons according to $f_{i}(T)=\max_{c}g_{i}(T,c)$,
where $c$ is the half difference between the strengths of the presented
stimuli (\textbf{eq. \ref{eq:StimStrengthDM}}).

\subsubsection{Temporal scaling in the production epoch}

Analysis of temporal scaling was performed using similar technique
to Ref. \cite{Wang_2018}. Specifically, we calculated the $k$th
scaling component $\mathbf{u}_{SC,k}$ through the following equation:
\begin{equation}
\mathbf{u}_{SC,k}=\text{arg }\min_{\mathbf{u}}\frac{\sum_{t}\sum_{T}(\mathbf{r}_{k}^{S}(t;T)\mathbf{u}-\text{Mean}_{T}(\mathbf{r}_{k}^{S}(t;T)\mathbf{u}))^{2}}{\sum_{t}\sum_{T}(\mathbf{r}_{k}^{S}(t;T)\mathbf{u}-\text{Mean}_{\{t,T\}}(\mathbf{r}_{k}^{S}(t;T)\mathbf{u}))^{2}},\label{eq:scaling_component}
\end{equation}
where $\mathbf{r}_{k}^{S}(t;T)$ is population activity at the scaled
time when the duration of the perception epoch is $T$ (see below
for details), the denominator is the total variance of the trajectories,
and the numerator is the variance that cannot be explained by temporal
scaling. To calculate the first scaling component $\mathbf{u}_{SC,1}$,
we set $\mathbf{r}_{1}^{S}(t;T)=\mathbf{r}^{PC}(tT_{p};T),$ with
$0\le t\le1$, where $\mathbf{r}^{PC}$ is the projection of the
population activity in the subspace spanned by the first 9 principal
components, and $T_{p}$ is the interval produced by the network
in the production epoch; then we minimized $\mathbf{u}$ in eq. \ref{eq:scaling_component}.
To calculate the second scaling component $\mathbf{u}_{SC,2}$, we
set $\mathbf{r}_{2}^{S}(t;T)=\mathbf{r}_{1}^{S}(t;T)-\mathbf{r}_{1}^{S}(t;T)\mathbf{u}_{SC,1}$,
and then minimized $\mathbf{u}$ in eq. \ref{eq:scaling_component}
in the subspace orthogonal to $\mathbf{u}_{SC,1}$. In this way,
we calculated all the 9 scaling components one by one.

Scaling index (SI) of a subspace $U$ was defined as 
\begin{equation}
\text{SI}=\frac{\sum_{t}\sum_{T}(\mathbf{r}_{1}^{S}(t;T)U-\text{Mean}_{T}(\mathbf{r}_{1}^{S}(t;T)U))^{2}}{\sum_{t}\sum_{T}(\mathbf{r}_{1}^{S}(t;T)U-\text{Mean}_{\{t,T\}}(\mathbf{r}_{1}^{S}(t;T)U))^{2}},\label{eq:SI}
\end{equation}
where $\mathbf{r}_{1}^{S}(t;T)U$ is the projection of the scaled
trajectory to the subspace $U$.

\subsubsection{The geometry of coding combination}

During the perception epoch of t-SR, the network state is quantified
by the time elapsed from the beginning of the epoch (temporal flow)
and the spatial information of the first pulse. At the end of the
delay epoch of t-SR, the network state is quantified by the time
interval between the first two pulses and the spatial information
of the first pulse. During the production epoch of t-SR, the network
state is quantified by temporal flow, time interval and spatial information.
Similar scenario also exists in t-DM, except that the non-temporal
information is the decision choice made by the network. In t-DM,
the decision choice $d$ depends on the sign of the half difference
$c$ between the strength of the presented two stimuli (\textbf{eq.
\ref{eq:StimStrengthDM}}), we defined $r_{i}(d=1,\{a\})=\langle r_{i}(c,\{a\})\rangle_{c>0}$
and $r_{i}(d=-1,\{a\})=\langle r_{i}(c,\{a\})\rangle_{c<0}$, where
$\{a\}$ indicates the other parameters than decision choice, and
used $r_{i}(d,\{a\})$ to do the following analysis. Together, during
the perception epoch and at the end of the delay epoch of t-SR and
t-DM, two variables are coded in the network state; during the production
epoch, three variables are coded in the network state. We used two
measurements to quantify the geometry of the coding combination of
multiple variables: (1) the angle between the first marginal principal
components and (2) the mixed variance \cite{Kobak_2016}, introduced
below.

Suppose the activity of the $i$th neuron $r_{i}(a,b)$ is a function
of two variables $a$ and $b$, with the mean of $r_{i}(a,b)$ being
subtracted so that $\langle r_{i}(a,b)\rangle_{a,b}=0$. The marginal
principal components (PCs) with respect to $a$ are the PCs of the
dot set $\{\langle r_{i}(a,b)\rangle_{b}\}_{i}$, and the marginal
PCs of $b$ are the PCs of $\{\langle r_{i}(a,b)\rangle_{a}\}_{i}$.
We quantified the coding orthogonality of $a$ and $b$ by calculating
the angle between the first marginal PCs of $a$ and $b$. The portions
of variance explained by $a$ and $b$ are respectively $p_{a}=\text{Var}_{i,a}(\{\langle r_{i}(a,b)\rangle_{b}\}_{i})/v_{tot}$
and $p_{b}=\text{Var}_{i,b}(\{\langle r_{i}(a,b)\rangle_{a}\}_{i})/v_{tot}$,
with the total variance $v_{tot}=\text{Var}_{i,a,b}(\{r_{i}(a,b)\}_{i})$.
The portion of mixed variance between $a$ and $b$ is $p_{a+b}=1-p_{a}-p_{b}$. 

In the case that the activity of the $i$th neuron $r_{i}(a,b,c)$
is a function of three variables, we also subtracted the mean of
$r_{i}(a,b,c)$ so that $\langle r_{i}(a,b,c)\rangle_{a,b,c}=0$.
The marginal PCs of $a$, $b$ and $c$ are respectively the PCs
of $\{\langle r_{i}(a,b,c)\rangle_{b,c}\}_{i}$, $\{\langle r_{i}(a,b,c)\rangle_{a,c}\}_{i}$
and $\{\langle r_{i}(a,b,c)\rangle_{a,b}\}_{i}$. The portions of
variance explained by these variables and their mixing were defined
as \cite{Kobak_2016}:
\[
p_{a}=\text{Var}_{i,a}(\{\langle r_{i}(a,b)\rangle_{b,c}\}_{i})/v_{tot}
\]
\[
p_{b}=\text{Var}_{i,b}(\{\langle r_{i}(a,b)\rangle_{a,c}\}_{i})/v_{tot}
\]
\[
p_{c}=\text{Var}_{i,c}(\{\langle r_{i}(a,b)\rangle_{a,b}\}_{i})/v_{tot}
\]
\[
p_{a+b}=\text{Var}_{i,a,b}(\{\langle r_{i}(a,b,c)-\langle r_{i}(a,b)\rangle_{b,c}-\langle r_{i}(a,b)\rangle_{a,c}-\langle r_{i}(a,b)\rangle_{a,b}\rangle_{c}\}_{i})/v_{tot}
\]
\[
p_{b+c}=\text{Var}_{i,b,c}(\{\langle r_{i}(a,b,c)-\langle r_{i}(a,b)\rangle_{b,c}-\langle r_{i}(a,b)\rangle_{a,c}-\langle r_{i}(a,b)\rangle_{a,b}\rangle_{a}\}_{i})/v_{tot}
\]
\[
p_{a+c}=\text{Var}_{i,a,c}(\{\langle r_{i}(a,b,c)-\langle r_{i}(a,b)\rangle_{b,c}-\langle r_{i}(a,b)\rangle_{a,c}-\langle r_{i}(a,b)\rangle_{a,b}\rangle_{b}\}_{i})/v_{tot}
\]
\[
p_{a+b+c}=1-p_{a}-p_{b}-p_{c}-p_{a+b}-p_{b+c}-p_{a+c}
\]
where $v_{tot}=\text{Var}_{i,a,b,c}(\{r_{i}(a,b,c)\}_{i})$ is the
total variance, ``$+$'' sign in the subscript indicates the mixing
of several variables. 

In \textbf{Fig. 3},\textbf{ }we used the network state trajectory
after 400 ms (200 ms) of transient period of the perception (production)
epoch to do the analysis.

\subsubsection{Decoding}

We studied two types of nearest-centroid decoders \cite{Murray_2017}.
Given a population state $\mathbf{f}_{0}$, the decoded value $a_{d,1}$
read-out by Decoder 1 is 
\begin{equation}
a_{d,1}=\text{arg min}_{a\in\mathcal{A}}(\left\Vert \mathbf{f}_{0}\mathbf{W}^{dec}-\mathbf{f}(a;b_{train})\mathbf{W}^{dec}\right\Vert ),\label{eq:Decoder1}
\end{equation}
where $\mathbf{f}(a;b_{train})$ is the population state as a function
of variable $a$ along an iso-$b$ line whose $b$ value is constantly
$b_{train}$, and decoding weight $\mathbf{W}^{dec}$ is the first
PC of $\mathbf{f}(a;b_{train})$. The decoded value $a_{d,2}$ read-out
by Decoder 2 is 
\begin{equation}
a_{d,2}=\text{arg min}_{a\in\mathcal{A}}(\left\Vert (\mathbf{f}_{0}-\langle\mathbf{f}(a;b_{test})\rangle_{a})\mathbf{W}^{dec}-(\mathbf{f}(a;b_{train})-\langle\mathbf{f}(a;b_{train})\rangle_{a})\mathbf{W}^{dec}\right\Vert ),\label{eq:Decoder2}
\end{equation}
where $\mathbf{f}(a;b_{test})$ is the iso-$b$ line that $\mathbf{f}_{0}$
belongs to, and $\langle\cdot\rangle_{a}$ means averaging over $a$.
From eq.\ref{eq:Decoder2}, both the mass centers of the two iso-$b$
lines $\mathbf{f}(a;b_{train})$ and $\mathbf{f}(a;b_{test})$ are
translationally moved to the zero point before $\mathbf{f}(a;b_{train})$
and $\mathbf{f}(a;b_{test})$ are projected to the decoding space
by $\mathbf{W}^{dec}$.

\subsubsection{Correlation between decoding error, angle and mixed variance\label{sub:Corr(AG,MV)}}

In \textbf{Fig. 4d, f}, we computed the correlation between decoding
error (DE), the angle (AG) between the first PCs of the decoded and
generalized variables, and the mixed variance (MV) between the decoded
and generalized variables. A subtle point here is that AG and MV
may also be correlated (see \textbf{Fig. 4c, e }for the negative
correlation between AG and MV in the production epoch of t-SR), therefore
the Pearson's correlation between DE and AG may be contributed by
two pathways: (1) AG influences DE directly; (2) AG influences DE
\textit{indirectly} through MV, due to the correlation between AG
and MV. Similar situation also exists for the correlation between
DE and MV. To investigate the direct correlation and remove the indirect
one, we iteratively took the following operation to reduce the correlation
between AG and MV: removing a single data point (i.e., the AG and
MV of a single training configuration) from the dataset, so that
the absolute value of the correlation between AG and MV in the left
dataset is minimal. We found that small correlation (with absolute
value below 0.05) between AG and MV could usually be obtained after
removing 2 or 3 data points from the whole dataset of 30 points (\textbf{Figs.
\ref{fig:Timed-spatial-reproduction-decoding}, \ref{fig:Timed-decision-making-decoding}}).
In this way, we got a dataset with small correlation between AG and
MV, while at the same time, as large as possible. Pearson's correlation
were then calculated using the left dataset to draw \textbf{Figs.
4d, f, \ref{fig:Timed-spatial-reproduction-decoding}, \ref{fig:Timed-decision-making-decoding}}.

\subsubsection{Firing sequence and network structure}

To plot \textbf{Fig. 5a, b}, we ordered the peak firing time of strongly
active neurons (whose peak firing rates were larger than 2) in the
studied epoch, and plotted weight connection as a function of the
peak order difference between the post- and pre-synaptic neurons.

To plot \textbf{Fig. 5c, d}, we used a more elaborate method to illustrate
the network structure underlying t-SR and t-DM. At time $t_{0}$
and non-time information $x_{0}$ (which may be spatial location
or decision choice), we picked a set $\mathcal{N}(t_{0},x_{0})$
of strongly active neurons whose firing rates at $t_{0}$ and $x_{0}$
were larger than a threshold 2 (our result is insensitive to this
threshold). We then defined\textit{ }$T_{peak,i}(t_{0},x_{0})$ to
be the peak time of neuron $i$ near $t_{0}$ at $x_{0}$: if the
activity $f_{i}(t;x_{0})$ of neuron $i$ decreased (or increased)
with time at time point $t_{0}$ and non-time information $x_{0}$,
then $T_{peak,i}(t_{0},x_{0})$ was the time point of the local maximum
of $f_{i}(t;x_{0})$ before (or after), but most nearest to, $t_{0}$.
Iterating over all the possible values of $x_{0}$, we got all the
strongly active neurons at time $t_{0}$: $\mathcal{N}(t_{0})=\bigcup_{x_{0}}\mathcal{N}(t_{0},x_{0})$.
For neuron $i$ in $\mathcal{N}(t_{0})$, we called its prefered
non-time information $x_{prefer}$ to be the value of $x_{0}$ that
maximized its peak firing rate: $x_{prefer}=\text{arg max}_{x_{0}}f_{i}(T_{peak,i}(t_{0},x_{0}),x_{0})$.
In this way, we classified all the neurons in $\mathcal{N}(t_{0})$
according to their non-time information preference: $\mathcal{N}(t_{0})=\bigcup_{x_{0}}\mathcal{N}_{prefer}(t_{0},x_{0})$,
with $\mathcal{N}_{prefer}(t_{0},x_{0})$ being the set of neurons
that prefer $x_{0}$ around time $t_{0}$. We then defined $T_{peak,i}(t_{0},x_{prefer})$
to be the \textit{big peak time} of neuron $i$ at time $t_{0}$.
Given a neuron $i$ and a set $\mathcal{N}_{prefer}(t_{0},x_{0})$
of neurons ($i$ may or may not belong to $\mathcal{N}_{prefer}(t_{0},x_{0})$),
we ordered their big peak times, and then investigated the recurrent
weight from $i$ to each neuron of $\mathcal{N}_{prefer}(t_{0},x_{0})$
(except $i$ itself if $i\in\mathcal{N}_{prefer}(t_{0},x_{0})$).
In this way, we studied the recurrent weight $w(o_{post}-o_{pre},|x_{post}-x_{pre}|)$
as a function of the difference $o_{post}-o_{pre}$ between the orders
of the big peak time of the post- and pre-synaptic neurons and the
difference $|x_{post}-x_{pre}|$ of their preferred non-time information.
\textbf{Fig. 5c,d} were plotted by averaging $w(o_{post}-o_{pre},|x_{post}-x_{pre}|)$
over $t_{0}$ and training configurations.

\section{The relationship between the low dimensionality of the attractor
in the delay epoch and the dominance of monotonic neurons\label{sec:Monotonic_neurons}}

\begin{figure}[tbph]
\centering \includegraphics[scale=0.7]{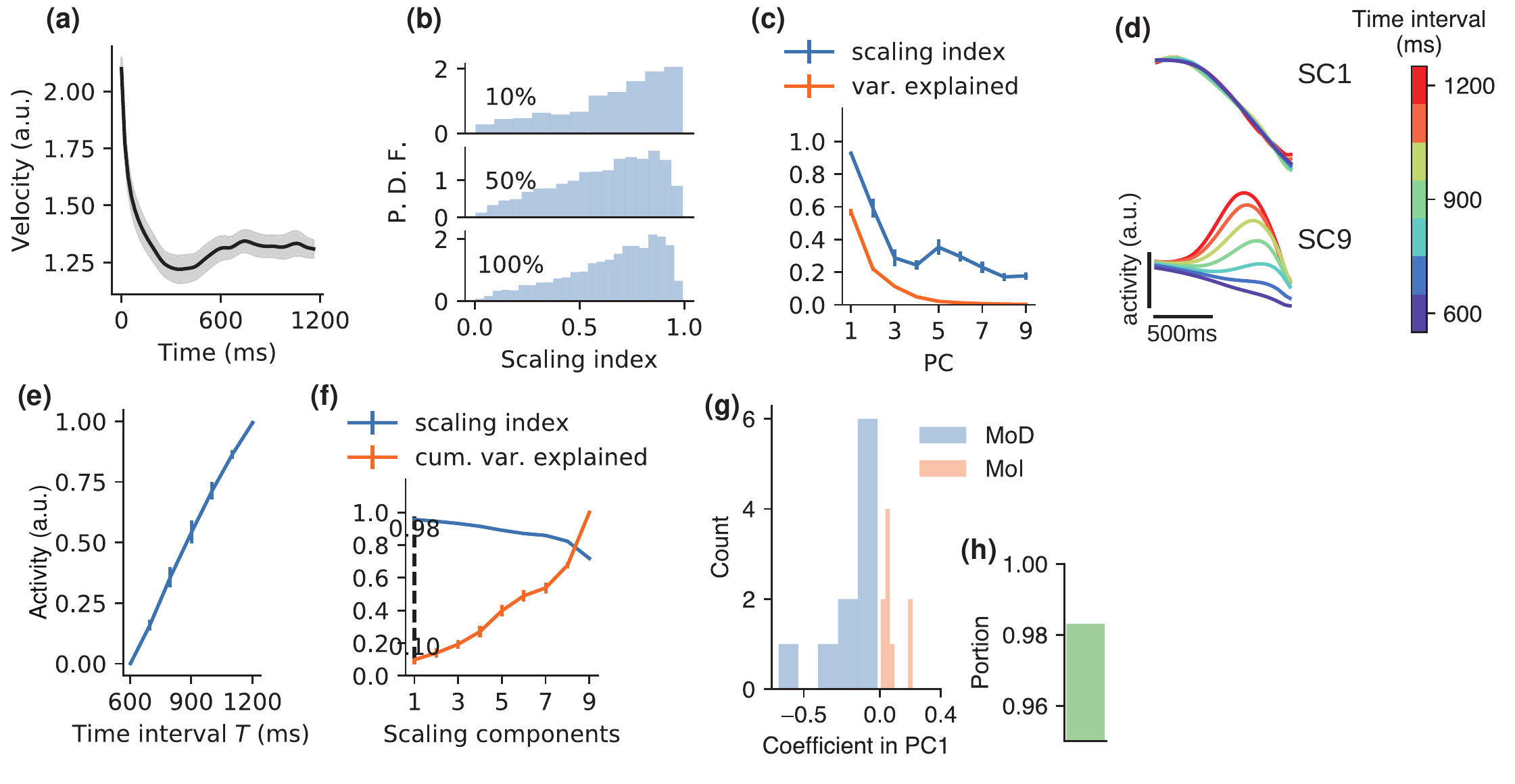}

\protect\protect\protect\caption{\textbf{\label{fig:Interval-production-task.}Interval production
task}. (\textbf{a}) Trajectory speed with time in the perception
epoch, shaded belt indicating s.e.m. (standard error of mean). (\textbf{b})
Probability distribution function (p.d.f) of scaling indexes of the activities of single neurons in the production epoch, after counting
neurons with the top 10\% highest activity (upper panel), top 50\%
(middle panel) and all neurons (lower panel). (\textbf{c}) The scaling
index and explained variance of principal components (PC) in the
production epoch. (\textbf{d}) We calculated the scaling components
in the subspace spanned by the first nine principal components. Shown
are the first (upper) and last (lower) scaling component of the production
epoch of an example training configuration. Color of lines indicate
to-be-produced interval $T$. (\textbf{e}) The mean activity of the
last scaling component as a function of $T$, with the activities when $T=600$ ms and $T=1200$ ms are respectively normalized to be 0 and 1. (\textbf{f}) Scaling index (blue) and ratio of explained variance (orange) in the subspace spanned by the accumulated scaling components. This panel is in the same style as \textbf{Fig. 2n}, except that it analyzes the perception epoch of IP task. (\textbf{g,h}) These
two panels explain the relationship between the low dimensionality
of manifold $\mathcal{M}$ at the end of the delay epoch and the
dominance of neurons monotonically tuned by $T$ (Section \ref{sec:Monotonic_neurons}).
(\textbf{g}) Histogram of the elements of PC1 of the manifold $\mathcal{M}$
at the end of the delay epoch at different $T$s of an example training
configuration. Note that the elements corresponding with monotonically
decreasing (MoD) and monotonically increasing (MoI) neurons have
different signs. (\textbf{h}) In 16 training configurations, for
a given element in PC1 of $\mathcal{M}$, it has over 98\% probability
to have the same sign with most other elements corresponding with
neurons of the same type, while have the opposite sign with most
other elements corresponding with neurons of the opposite type. In
panels \textbf{c,e}, error bars indicate s.e.m. over training configurations.}
\end{figure}

We denote $\mathcal{M}$ as the manifold of the population states
at the end of the delay epoch at different durations $T$ of the
perception epoch (\textbf{Fig. 2e}). The first principal component
(PC) of $\mathcal{M}$ explained about 90\% of its variance (\textbf{Fig.
2g}), and the activities of most neurons changed monotonically with
$T$ in $\mathcal{M}$ (\textbf{Fig. 2j}). To understand the relationship
between these two facts, let's consider the extreme case that all
neurons are linearly monotonic with $T$ in $\mathcal{M}$, then
$\mathcal{M}$ is a line in the population-state space that can be
parameterized as $[f_{1}(T),f_{2}(T),\cdots,f_{N}(T)]^{T}$, with
$f_{i}(T)$ being the activity of the $i$th neuron at the end of
the delay epoch when the duration of the perception epoch is $T$.
In this case, PC1 of $\mathcal{M}$, which explains 100\% of the
variance of $\mathcal{M}$ because $\mathcal{M}$ is a line, is the
following vector with unit length: 
\[
\pm\frac{1}{\sqrt{\sum_{i}(f_{i}(T_{max})-f_{i}(T_{min}))^{2}}}[f_{1}(T_{max})-f_{1}(T_{min}),f_{2}(T_{max})-f_{2}(T_{min}),\cdots,f_{N}(T_{max})-f_{N}(T_{min})]^{T},
\]
where $T_{min}=600\text{ms}$ and $T_{max}=1200\text{ms}$ are respectively
the minimal and maximal values of $T$ in our simulation, and the
$\pm$ sign indicates that the direction of PC1 is undetermined.
If neuron $i$ monotonically increases (or decreases) with $T$,
then $f_{i}(T_{max})-f_{i}(T_{min})>0$ (or $f_{i}(T_{max})-f_{i}(T_{min})<0$).
Apparently, if two neurons $i$ and $j$ have the same (or different)
monotonicity, then their corresponding elements in PC1 have the same
(different) signs. This is indeed what we found in our simulation
(\textbf{Fig. \ref{fig:Interval-production-task.}g, h}).

\section{The geometric meaning of mixed variance}

We denote the population state to be $\mathbf{r}=\{r_{1},r_{2},\cdots,r_{N}\}$,
where $r_{i}$ is the firing rate of the $i$th neuron, or in general,
the activity projected on the $i$th basis vector, say, principal
component. Suppose $\mathbf{r}$ is parameterized by two variables
$a$ and $b$, and we subtract the mean value of $r_{i}$ so that
\begin{equation}
\text{E}_{a,b}[r_{i}(a,b)]=0,\label{eq:mv_1}
\end{equation}
where $\text{E}_{a,b}[\cdot]$ means the average over $a$ and $b$. 

The total variance of $\mathbf{r}$ is 
\[
v_{tot}=\text{Var}_{i,a,b}[r_{i}(a,b)]
\]
\[
=\text{E}_{i}[\text{Var}_{a,b}[r_{i}(a,b)]]+\text{Var}_{i}[\text{E}_{a,b}[r_{i}(a,b)]]
\]
\begin{equation}
=\text{E}_{i}[\text{Var}_{a,b}[r_{i}(a,b)]],\label{eq:mv_2}
\end{equation}
where $\text{Var}_{x}[\cdot]$ means the variance over variable $x$.
The first equation is the definition of the total variance, the second
equation is from the law of total variance, and the third equation
is from eq. \ref{eq:mv_1}. Similarly, the variance explained by
$a$ is 
\begin{equation}
v_{a}=\text{Var}_{i,a}[\text{E}_{b}[r_{i}(a,b)]]=\text{E}_{i}[\text{Var}_{a}[\text{E}_{b}[r_{i}(a,b)]]],\label{eq:mv_3}
\end{equation}
and the variance explained by $b$ is 
\begin{equation}
v_{b}=\text{Var}_{i,b}[\text{E}_{a}[r_{i}(a,b)]]=\text{E}_{i}[\text{Var}_{b}[\text{E}_{a}[r_{i}(a,b)]]]\label{eq:mv_4}
\end{equation}

Now let's study a sufficient condition so that 
\begin{equation}
v_{tot}=v_{a}+v_{b},\label{eq:mv_5}
\end{equation}
which means that the mixed variance 
\begin{equation}
v_{mix}=v_{tot}-(v_{a}+v_{b})\label{eq:mv_mix_variance}
\end{equation}
is zero.

From eqs. \ref{eq:mv_2}-\ref{eq:mv_4}, a sufficient condition to
fulfill eq. \ref{eq:mv_5} is 
\begin{equation}
\text{Var}_{a,b}[r_{i}(a,b)]=\text{Var}_{a}[\text{E}_{b}[r_{i}(a,b)]]+\text{Var}_{b}[\text{E}_{a}[r_{i}(a,b)]]\quad\text{for every }i.\label{eq:mv_6}
\end{equation}
According to the law of total variance, 
\begin{equation}
\text{Var}_{a,b}[r_{i}(a,b)]=\text{Var}_{a}[\text{E}_{b}[r_{i}(a,b)]]+\text{E}_{a}[\text{Var}_{b}[r_{i}(a,b)]].\label{eq:mv_7}
\end{equation}
Therefore, to realize eq. \ref{eq:mv_6}, we can set 
\begin{equation}
\text{Var}_{b}[\text{E}_{a}[r_{i}(a,b)]]=\text{E}_{a}[\text{Var}_{b}[r_{i}(a,b)]]\quad\text{for every }i.\label{eq:mv_8}
\end{equation}
in other words
\begin{equation}
\text{E}_{b}[(\text{E}_{a}[r_{i}(a,b)]-\text{E}_{a,b}[r_{i}(a,b)])^{2}]=\text{E}_{a}[\text{E}_{b}[(r_{i}(a,b)-\text{E}_{b}[r_{i}(a,b)])^{2}]]\quad\text{for every }i\label{eq:mv_9}
\end{equation}
Because $\text{E}_{a,b}[r_{i}(a,b)]=0$, this equation gives 
\begin{equation}
\text{E}_{b}[(\text{E}_{a}[r_{i}(a,b)])^{2}]=\text{E}_{a}[\text{E}_{b}[(r_{i}(a,b)-\text{E}_{b}[r_{i}(a,b)])^{2}]]\quad\text{for every }i\label{eq:mv_10}
\end{equation}
A sufficient condition to fulfill the equation above is 
\begin{equation}
r_{i}(a,b)-\text{E}_{b}[r_{i}(a,b)]=f(b)\quad\text{for every }i,\label{eq:mv_11}
\end{equation}
namely the value of $r_{i}(a,b)-\text{E}_{b}[r_{i}(a,b)]$ does not
depend on $a$. This sufficient condition can be easily proved by substituting eq. \ref{eq:mv_11}
into eq. \ref{eq:mv_10} and using the fact that $\text{E}_{a,b}[r_{i}(a,b)]=0$.
Now let's try to understand the meaning of eq. \ref{eq:mv_11}. Consider
four pairs of variables $(a_{1},b_{1})$, $(a_{2},b_{1})$, $(a_{1},b_{2})$
and $(a_{2},b_{2})$, we have 
\begin{equation}
r_{i}(a_{1},b_{1})-\text{E}_{b}[r_{i}(a_{1},b)]=f(b_{1})=r_{i}(a_{2},b_{1})-\text{E}_{b}[r_{i}(a_{2},b)]\quad\text{for every }i\label{eq:mv_12}
\end{equation}
\begin{equation}
r_{i}(a_{1},b_{2})-\text{E}_{b}[r_{i}(a_{1},b)]=f(b_{2})=r_{i}(a_{2},b_{2})-\text{E}_{b}[r_{i}(a_{2},b)]\quad\text{for every }i\label{eq:mv_13}
\end{equation}
By subtracting eq. \ref{eq:mv_12} from eq. \ref{eq:mv_13}, we have
\begin{equation}
r_{i}(a_{1},b_{1})-r_{i}(a_{1},b_{2})=r_{i}(a_{2},b_{1})-r_{i}(a_{2},b_{2})\quad\text{for every }i.\label{eq:mv_14}
\end{equation}
This means that between the two iso-$b$ lines in which the values of
$b$ are separately fixed at $b_{1}$ and $b_{2}$, the vector that
connects the two points representing $a_{1}$ is equal to the
vector that connects the two points representing $a_{2}$. In other
words, these two iso-$b$ lines can be related by translational movement.
By rewritten eq.\ref{eq:mv_14} as $r_{i}(a_{1},b_{1})-r_{i}(a_{2},b_{1})=r_{i}(a_{1},b_{2})-r_{i}(a_{2},b_{2})$,
we see that different iso-$a$ lines are also related by translational
movement. 

From the discussion above, translational relation between different
iso-$a$ or iso-$b$ lines is a sufficient condition for zero mixed
variance. How about the necessity? In other words, if we observe
close-to-zero mixed variance in simulation, how will be the geometry
of the iso-$a$ and iso-$b$ lines? We checked this point through
simulation. In \textbf{Fig. \ref{fig:Grid-geometry-examples}}, we
show the iso-space lines of several simulation examples, in the
perception, delay and production epochs of t-SR task. We see that in examples
with small mixed variance, the iso-space lines of different spatial
information tend to be parallel and of the same length; whereas in examples
with large mixed variance, the iso-space lines may be non-parallel
or of very different lengths. Additionally, if iso-$a$ or iso-$b$
lines are translationally related, then Decoder 2 (\textbf{eq. \ref{eq:Decoder2}})
will have perfectly zero generalization error. We found that the
generalization error of Decoder 2 is strongly positively correlated
with mixed variance (\textbf{Figs. 4f, \ref{fig:Timed-spatial-reproduction-decoding},
\ref{fig:Timed-decision-making-decoding}}). These results imply
that at least in the context of our simulation, mixed variance is
a good index to quantify the translational relationship between different
iso-$a$ or iso-$b$ lines, or in other words, the parallelogram-likeness
of iso-$a$ and iso-$b$ grids (\textbf{Fig. 3f, upper left}).

The opposite extreme case that $v_{mix}=v_{tot}$, which, from eq.\ref{eq:mv_mix_variance},
means $v_{a}=v_{b}=0$. From eqs. \ref{eq:mv_3}, \ref{eq:mv_4},
this means that 
\[
\text{Var}_{a}[\text{E}_{b}[r_{i}(a,b)]]=\text{Var}_{b}[\text{E}_{a}[r_{i}(a,b)]]=0\quad\text{for every }i.
\]
In other words, the mean value of $r_{i}(a,b)$ over $b$ (i.e., $\text{E}_{b}[r_{i}(a,b)]$) does not
depends on $a$, and the mean value of $r_{i}(a,b)$ over $a$ (i.e., $\text{E}_{a}[r_{i}(a,b)]$) does
not depends on $b$ neither. This implies that different iso-$a$
(and also iso-$b$) lines are strongly intertwined with each other,
so that they have the same mean state value. A good example of this
case is that every point in the 2-dimensional range of variables
$[a_{min},a_{max}]\otimes[b_{min},b_{max}]$ (where $a_{min}$, $a_{max}$,
$b_{min}$ and $b_{max}$ are the minimal and maximal values of $a$
and $b$ respectively) is mapped toward a random point in a state
space $[r_{1,min},r_{1,max}]\otimes[r_{2,min},r_{2,max}]\otimes\cdots\otimes[r_{n,min},r_{n,max}]$:
in this case, every iso-$a$ or iso-$b$ dot set of states has
the mean value located at the center of the state space $(\frac{r_{1,min}+r_{1,max}}{2},\frac{r_{2min}+r_{2,max}}{2},\cdots,\frac{r_{n,min}+r_{n,max}}{2})$. 

\begin{figure}[tbph]
\centering \includegraphics[scale=0.65]{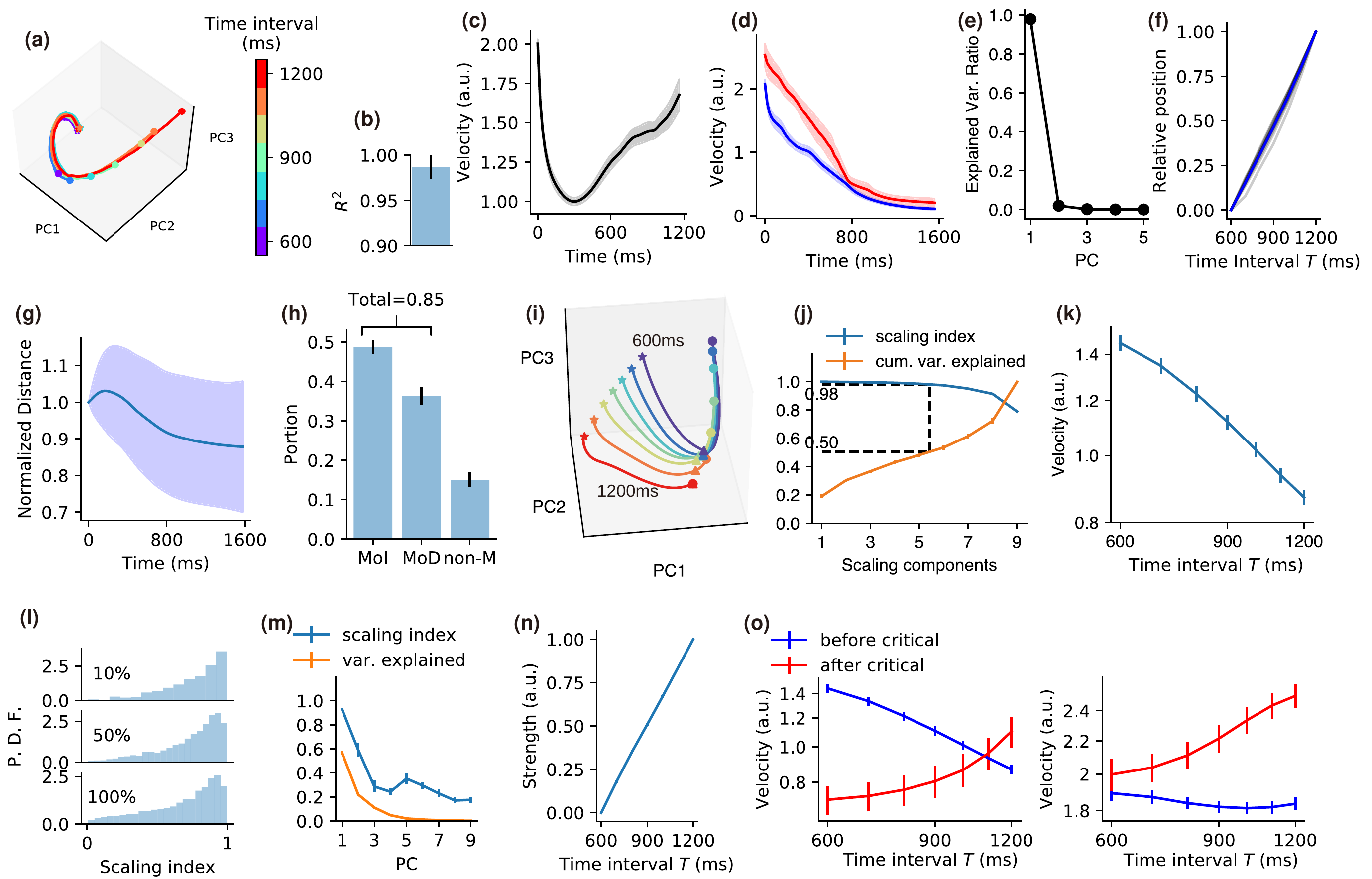}

\protect\protect\protect\caption{\textbf{\label{fig:Interval-comparison-task}Interval comparison
tasks}. (\textbf{a-c}) Stimulus1 epoch. (\textbf{a}) Population activity
in the stimulus1 epoch in the subspace of the first three PCs. Colors
indicate the duration $T$ of the epoch. Stars and circles respectively
indicate the starting and ending points of the stimulus1 epoch. (\textbf{b})
Coefficient of determination ($R^{2}$) that quantifies the overlap
of the firing profiles of individual neurons at different $T$s,
in the same style as \textbf{Fig. 2d} in the main text. (\textbf{c})
Trajectory speed as a function of time in the stimulus1 epoch, shaded
belt indicating s.e.m. (\textbf{d-h}) Delay epoch. (\textbf{d}) Trajectory
speed in the delay epoch when $T=600$ ms (blue) and 1200 ms (red),
in the same style as \textbf{Fig. 2f}. (\textbf{e}) Ratio of explained
variance of the first five PCs of manifold $\mathcal{M}$ at the
end of the delay epoch, in the same style as \textbf{Fig. 2g}. (\textbf{f})
The position of the state at the end of the delay epoch projected
in the first PC of manifold $\mathcal{M}$ as a function of $T$,
in the same style as \textbf{Fig. 2h}. (\textbf{g}) The distance
between two adjacent curves in the delay epoch as a function of time,
in the same style as \textbf{Fig. 2i}. (\textbf{h}) The portions
of monotonically decreasing (MoD), monotonically increasing (MoI),
and non-monotonic (non-M) types of neurons at the end of the delay
epoch, in the same style as \textbf{Fig. 2k}. (\textbf{j-o}) Stimulus2
epoch. (\textbf{i}) Population activity in the stimulus2 epoch in
the subspace of the first three PCs. The meanings of color scheme,
stars and circles are the same as panel \textbf{a}. Triangles indicate
critical points. The duration of stimulus 2 is kept at 1200 ms. (\textbf{j})
Scaling index (blue) and ratio of explained variance (orange) in
the subspace spanned by the accumulated scaling components, in the
same style as \textbf{Fig. 2n}. In this panel and panels \textbf{k-n},
only the trajectories from the beginning of stimulus 2 to the critical
points are studied. (\textbf{k}) Trajectory speed in the subspace
of the first three scaling components, in the same style as \textbf{Fig.
2o}. (\textbf{l}) Probability distribution of the scaling indexes
of single neurons, in the same style as \textbf{Fig. \ref{fig:Interval-production-task.}b}.
(\textbf{m}) The scaling index and explained variance of principal
components, in the same style as \textbf{Fig. \ref{fig:Interval-production-task.}c}.
(\textbf{n}) Mean activity of the last scaling component as a function
of $T$, in the same style as \textbf{Fig. \ref{fig:Interval-production-task.}e}.
(\textbf{o}) Left panel: speed of the trajectory before (blue) and
after (red) the critical point in the subspace of the first three
scaling components (SC). SCs are calculated using the trajectories
before the cirtical points, the red line is plotted by projecting
the trajectories after the critical points into the subspace of SCs
calculated using those before critical points. Right panel: speed
of the trajectory before (blue) and after (red) the critical point
in the full population state space. }
\end{figure}

\begin{figure}[tbph]
\centering \includegraphics[scale=0.7]{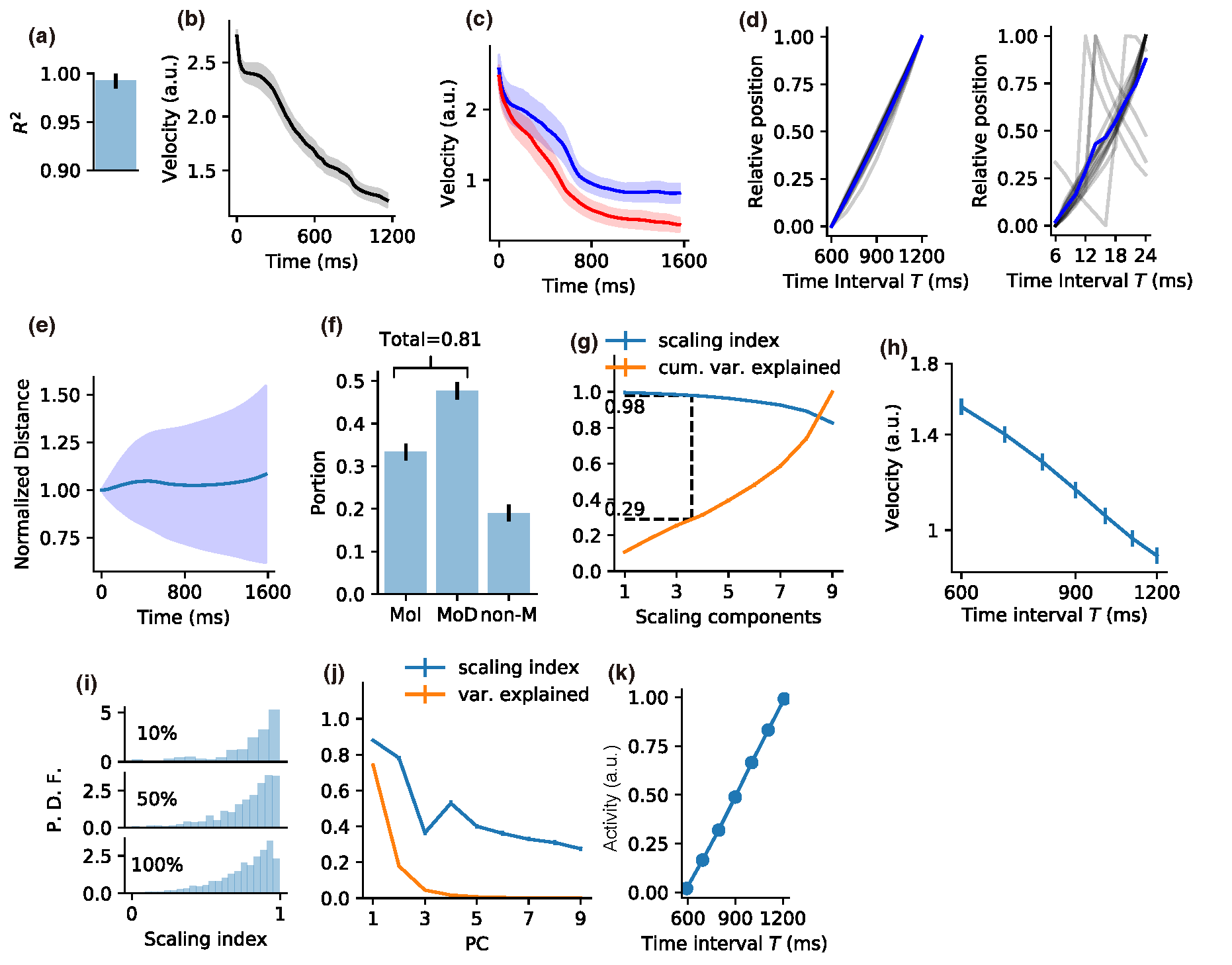}

\protect\protect\protect\caption{\textbf{\label{fig:Timed-spatial-reproduction}Timed spatial reproduction
task}. (\textbf{a,b}) Perception epoch. (\textbf{a}) Coefficient of determination ($R^{2}$) that
quantifies the overlap of the firing profiles of individual neurons
at different $T$s in the perception epoch, in the same style as
\textbf{Fig. 2d}. (\textbf{b}) Trajectory speed as a function of time in the perception epoch, shaded
belt indicating s.e.m. (\textbf{c-f}) Delay epoch. (\textbf{c}) Trajectory
speed as a function of time in the delay epoch when $T=600$ ms (blue)
and 1200 ms (red), in the same style as \textbf{Fig. 2f}. (\textbf{d})
The manifold $\mathcal{M}$ at the end of the delay epoch are parameterized
by both time interval $T$ between the first two pulses and the spatial
location $x$ of the first pulse. We denote $\mathcal{M}(T;x_{0})$
(or $\mathcal{M}(x;T_{0})$) to be the set of dots in $\mathcal{M}$
at specific location $x_{0}$ (or time interval $T_{0}$). Left panel:
the position of the state at the end of the delay epoch projected
to the first PC of $\mathcal{M}(T;x_{0})$ as a function of $T$,
with the position when $T=600$ ms (or 1200 ms) normalized to be
0 (or 1), in the same style as \textbf{Fig. 2h}. Gray curves: results
from 16 training configurations, each at a randomly chosen $x_{0}$.
Blue curve: mean value averaging over $x_{0}$ and training configurations.
Right panel: the position of the state in the first PC of $\mathcal{M}(x;T_{0})$.
We see that in most training configurations, the position in $\mathcal{M}(x;T_{0})$
encodes $x$ continuously and linearly, but big jump happens in some
configurations. (\textbf{e}) The distance between two adjacent curves
in the delay epoch as a function of time, similar to \textbf{Fig.
2i}. (\textbf{f}) The portions of monotonically decreasing (MoD),
monotonically increasing (MoI) and non-monotonic (non-M) types of
neurons tuned by $T$ at the end of the delay epoch, in the same
style as \textbf{Fig. 2k}. (\textbf{g-k}) Production epoch. (\textbf{g}) Scaling
index (blue) and ratio of explained variance (orange) in the subspace
spanned by the accumulated scaling components in the production epoch,
averaging over spatial locations and training configurations, in
the same style as \textbf{Fig. 2n}. (\textbf{h}) Trajectory speed
in the subspace of the first three scaling components in production
epoch, in the same style as \textbf{Fig. 2o}. (\textbf{i}) Probability
distribution of the scaling indexes of single neurons, in the same
style as \textbf{Fig. \ref{fig:Interval-production-task.}b}. (\textbf{j})
The scaling index and explained variance of principal components,
similar to \textbf{Fig. \ref{fig:Interval-production-task.}c}. (\textbf{k})
Mean activity of the last scaling component, similar to \textbf{Fig.
\ref{fig:Interval-production-task.}e}. Error bars representing s.e.m. are much smaller than the plot markers. }
\end{figure}

\begin{figure}[tbph]
\centering \includegraphics[scale=0.65]{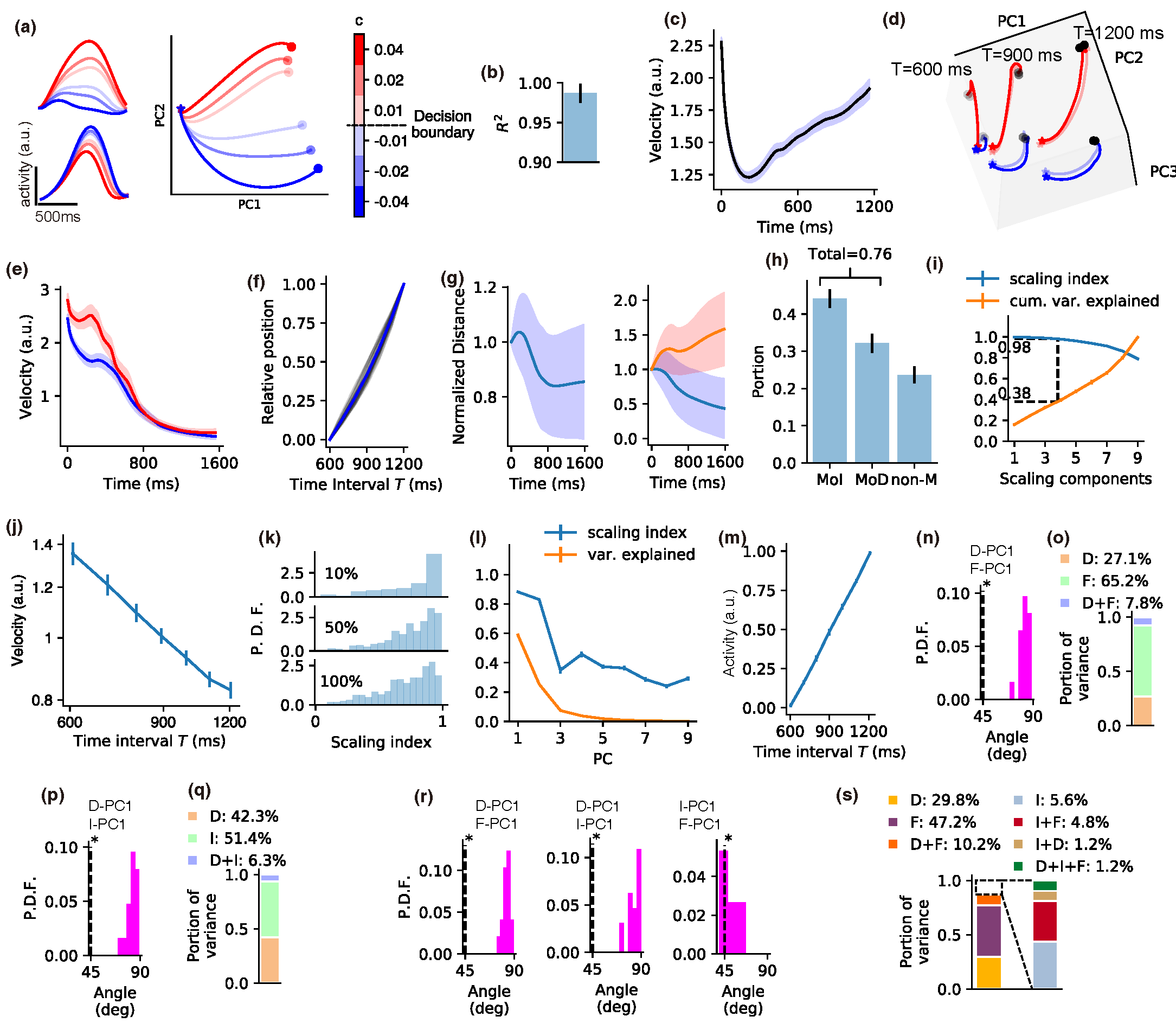}

\protect\protect\protect\caption{\textbf{\label{fig:Timed-decision-making}Timed decision making task}.
(\textbf{a-c}) Perception epoch. (\textbf{a}) Left: Firing profiles
of two example neurons in the perception epoch. Colors indicate $c$
value, which is the half difference between the strength of the presented
stimuli. Right: Trajectories in the subspace of the first two PCs.
Stars and circles respectively indicate the starting and ending points
of the perception epoch. (\textbf{b}) Coefficient of determination
($R^{2}$) that quantifies the overlap of the firing profiles of
individual neurons at different $T$s in the perception epoch, in
the same style as \textbf{Fig. 2d}.  (\textbf{c}) Trajectory speed as a function of time in the perception epoch, shaded
belt indicating s.e.m. (\textbf{d-h}) Delay epoch. (\textbf{d})
Trajectories in the subspace of the first three PCs. Stars and circles
respectively indicate the starting and ending points of the delay
epoch. Blackness of circles indicates $T$ value as annotated. Curve
color indicates $c$ value as indicated in the color map of panel
\textbf{a}, only $c=-0.04,$ -0.01, 0.01, 0.04 cases are plotted.
(\textbf{e}) Trajectory speed as a function of time in the delay
epoch when $T=600$ ms (blue) and 1200 ms (red), in the same style
as \textbf{Fig. 2f}. (\textbf{f}) The position of the state in the
first PC of $\mathcal{M}(T;d_{0})$ as a function of $T$, with the
position when $T=600$ ms (or 1200 ms) normalized to be 0 (or 1),
in the same style as \textbf{Fig. 2h}. Here, $\mathcal{M}(T;d_{0})$
represents the set of dots in manifold $\mathcal{M}$ at the end
of the delay epoch at specific decision choice $d_{0}$. (\textbf{g})
The distance between two adjacent curves in the delay epoch as a
function of time, in a similar style to \textbf{Fig. 2i}. Left panel:
the two adjacent curves have the same $c$ value, but slightly different
$T$ values. Right panel: the two adjacent curves have the same $T$
value, but different $c$ values. In the right panel, blue (orange)
curve represents the case when their $c$ values have the same (different)
sign, so that they have the same (different) decision choice. We
see that two trajectories representing the same (different) choice
tend to get close to (far away from) each other, consistent with
the scenario in panel \textbf{d}. (\textbf{h}) The portions of monotonically
decreasing (MoD), monotonically increasing (MoI) and non-monotonic
(non-M) types of neurons tuned by $T$ at the end of the delay epoch,
in the same style as \textbf{Fig. 2k}. (\textbf{i-m}) Production
epoch. (\textbf{i}) Scaling index (blue) and ratio of explained variance
(orange) in the subspace spanned by the accumulated scaling components,
averaging over $c$ values and training configurations, in the same
style as \textbf{Fig. 2n}. (\textbf{j}) Trajectory speed in the subspace
of the first three scaling components, in the same style as \textbf{Fig.
2o}. (\textbf{k}) Probability distribution of the scaling indexes
of single neurons, in the same style as \textbf{Fig. \ref{fig:Interval-production-task.}b}.
(\textbf{l}) The scaling index and explained variance of principal
components, in the same style as \textbf{Fig. \ref{fig:Interval-production-task.}c}.
(\textbf{m}) Mean activity of the last scaling component, in the
same style as \textbf{Fig. \ref{fig:Interval-production-task.}e}.
(\textbf{n-s}) The angle between first parameter-marginalized principal
components and mixed variances in the perception (panels \textbf{n,o}),
delay (panels \textbf{p,q}) and production epochs (panels \textbf{r,s}).
These panels are in the same style as \textbf{Fig. 3d, e, g-j}, except
that the non-spatial information is decision choice. }
\end{figure}

\begin{figure}[tbph]
\centering \includegraphics[scale=0.7]{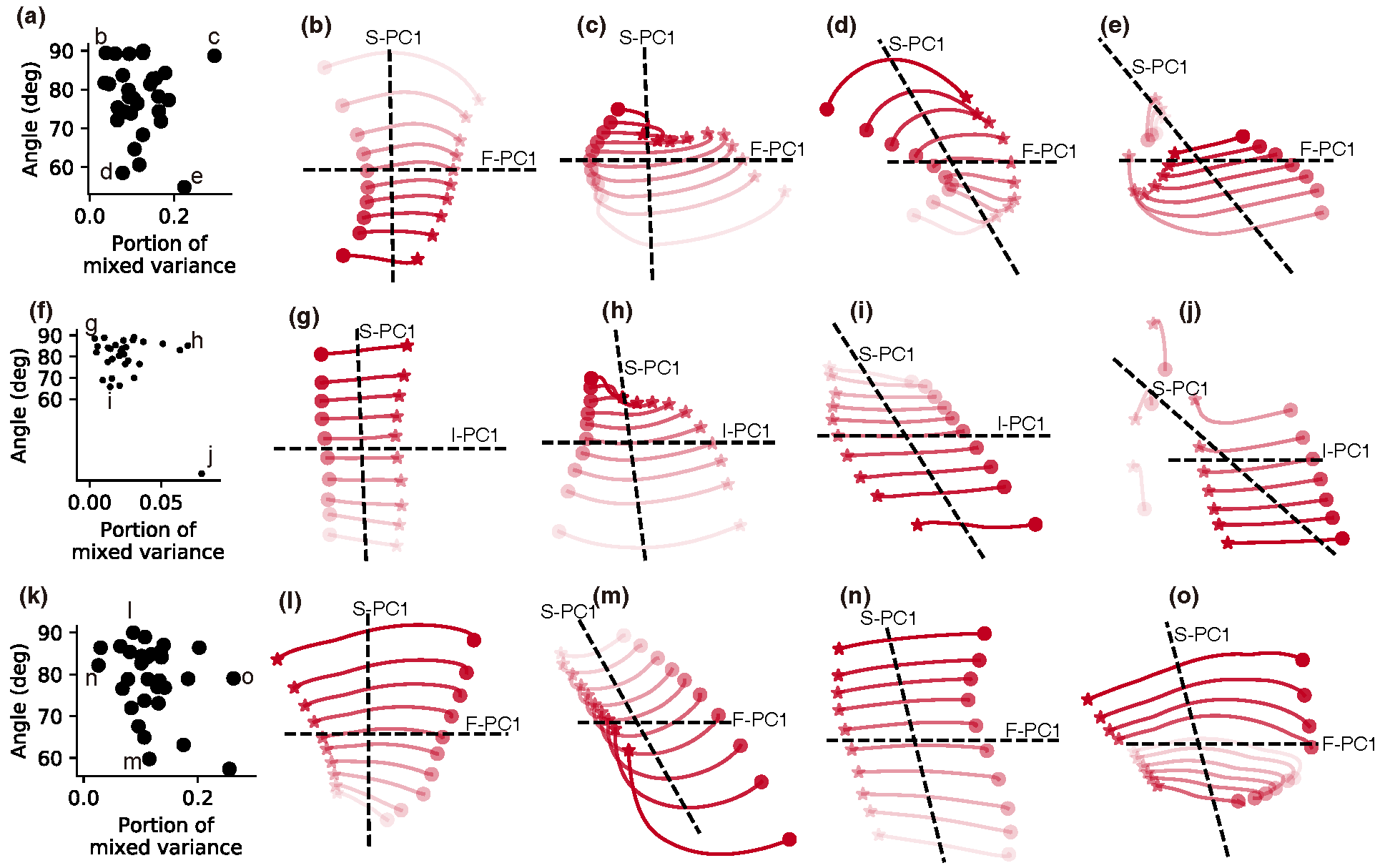}

\protect\protect\protect\caption{\textbf{\label{fig:Grid-geometry-examples}Examples that illustrate
the geometry of the coding combination of temporal and spatial information
in t-SR. }(\textbf{a-e}) Perception epoch. (\textbf{a}) Each dot
represents the angle between F-PC1 and S-PC1 as well as their mixed
variances in the perception epoch (after 400 ms of transient period)
of t-SR in a training configuration. (\textbf{b-e}) Iso-space lines
in the subspace spanned by F-PC1 and S-PC1, in the training configurations
indicated in panel \textbf{a}.\textbf{ }Stars indicate the points
after 400 ms of transient period from the beginning of the perception
epoch, and circles indicate the ending points of the perception epoch.
Redness from light to strong indicates the spatial locations $x=0,2,4,\cdots,18$.
(\textbf{f-j}) The same as panels \textbf{a-e}, except for showing
the iso-space lines in the manifold $\mathcal{M}$ at the end of
the delay epoch, in the subspace spanned by the first time-interval
PC (I-PC1) and S-PC1. Stars and circles indicate $T=600$ ms and
1200 ms cases respectively. (\textbf{k-o}) The same as panels \textbf{a-e},
except that the iso-space lines in the production epoch are shown. Stars indicate
the points after 200 ms of transient period from the beginning of
the production epoch, and circles indicate the ending points of the
production epoch. }
\end{figure}

\begin{figure}[tbph]
\centering \includegraphics[scale=0.65]{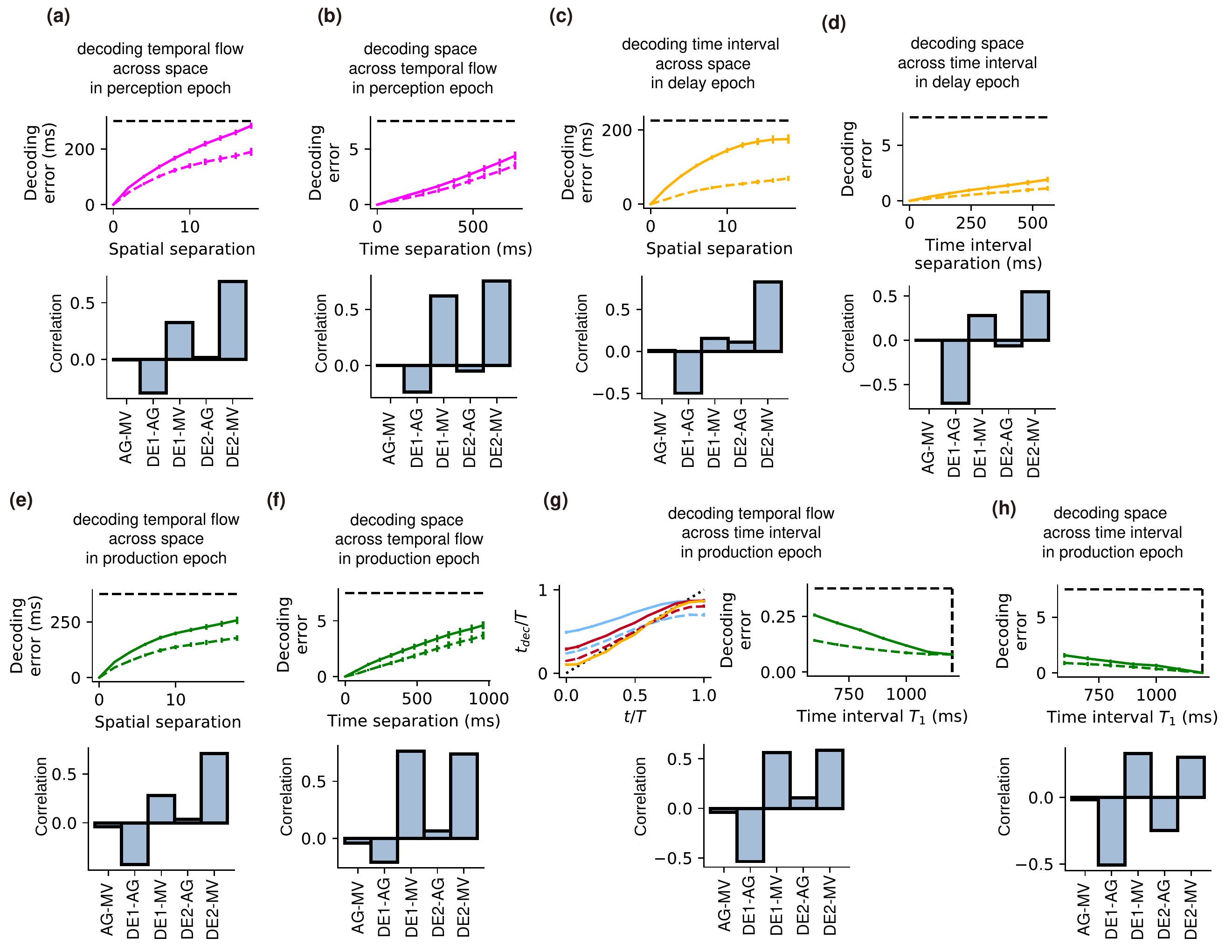}

\protect\protect\protect\caption{\textbf{\label{fig:Timed-spatial-reproduction-decoding}Decoding
generalizability in t-SR.} (\textbf{a-b}) Perception epoch. (\textbf{a})
Upper: decoding error as a function of $|x_{train}-x_{test}|$, after
Decoder 1 (solid line) or Decoder 2 (dashed line) is trained to read
the time elapsed from the beginning of the perception epoch (i.e.,
temporal flow) using the state trajectory at spatial location $x_{train}$,
and then tested at spatial location $x_{test}$, in the same style
as \textbf{Fig. 4g}. Horizontal dashed line indicates chance level,
supposing the decorder works by random guess. Lower: The correlations
between the angle (AG) between the first temporal-flow PC and the
first spatial PC, the mixed variance (MV) between temporal flow and
spatial information, the error of Decoder 1 (DE1) and the error of
Decoder 2 (DE2), in the same style as \textbf{Fig. 4d, f}. Note that
the correlation between AG and MV is approximately zero, see \textbf{Section
\ref{sub:Corr(AG,MV)}} for this point.  (\textbf{b}) Upper: Decoding
error as a function of $|t_{train}-t_{test}|$, after Decoder 1 (solid
line) or Decoder 2 (dashed line) is trained to read the spatial location
at time $t_{train}$ after the beginning of the perception epoch,
and then tested at time $t_{test}$. Lower: Correlations between AG,
MV, DE1 and DE2. (\textbf{c-d}) Delay epoch. (\textbf{c}) Similar
to panel \textbf{a}, except for decoding time interval across spatial
information using the state in manifold $\mathcal{M}$ at the end
of the delay epoch. (\textbf{d}) Decoding spatial information across
time interval using the states in manifold $\mathcal{M}$ at the
end of the delay epoch. (\textbf{e-h}) Production epoch. (\textbf{e})
Decoding temporal flow across spatial information in the production
epoch. The decoder was trained using $\mathbf{r}(t;x_{train},T_{0})$
and tested using $\mathbf{r}(t;x_{test},T_{0})$, where $\mathbf{r}(t;x_{0},T_{0})$
represents the population activity as a function of $t$ at specific
spatial information $x_{0}$ and time interval $T_{0}$. $T_{0}=1200$
ms in this panel and panels \textbf{f}. (\textbf{f}) Decoding space
across temporal flow in the production epoch. The decoder was trained
using $\mathbf{r}(x;t_{train},T_{0})$ and tested using $\mathbf{r}(x;t_{test},T_{0})$,
where $\mathbf{r}(x;t_{0},T_{0})$ represents the population activity
as a function of spatial information $x$ at specific time point
$t_{0}$ and time interval $T_{0}$. (\textbf{g}) Decoding temporal
flow across time interval in the production epoch. The decoder was
trained using $\mathbf{r}(t;T_{train},x_{0})$ and tested using $\mathbf{r}(t;T_{test},x_{0})$.
The results are averaged over $x_{0}\in[0,20]$. Upper left: The
decoded value $t_{dec}$ as a function of the time $t$ elapsed from
the beginning of the production epoch, after Decoder 1 (solid line)
or Decoder 2 (dashed line) was trained to read $t$ at $T=1200$
ms, and then tested at $T=600$ ms (blue), 900 ms (red) and 1200ms
(yellow). The dashed line indicates perfect temporal scaling. Upper
right: Decoding error as a function of $T$, after a decoder is trained
to read scaled temporal flow $t/T$ at $T=1200$ ms (indicated by
the vertical dashed line), and then tested at $T=T_{1}$. Lower:
correlations. (\textbf{h}) Decoding space across time interval in
the production epoch. The decoder was trained using $\langle\mathbf{r}(x;T_{train},t_{0})\rangle_{t_{0}}$
and tested using $\langle\mathbf{r}(x;T_{test},t_{0})\rangle_{t_{0}}$,
where $\langle\cdot\rangle_{t_{0}}$ means averaging over temporal
flow $t_{0}$. }
\end{figure}

\begin{figure}[tbph]
\centering \includegraphics[scale=0.65]{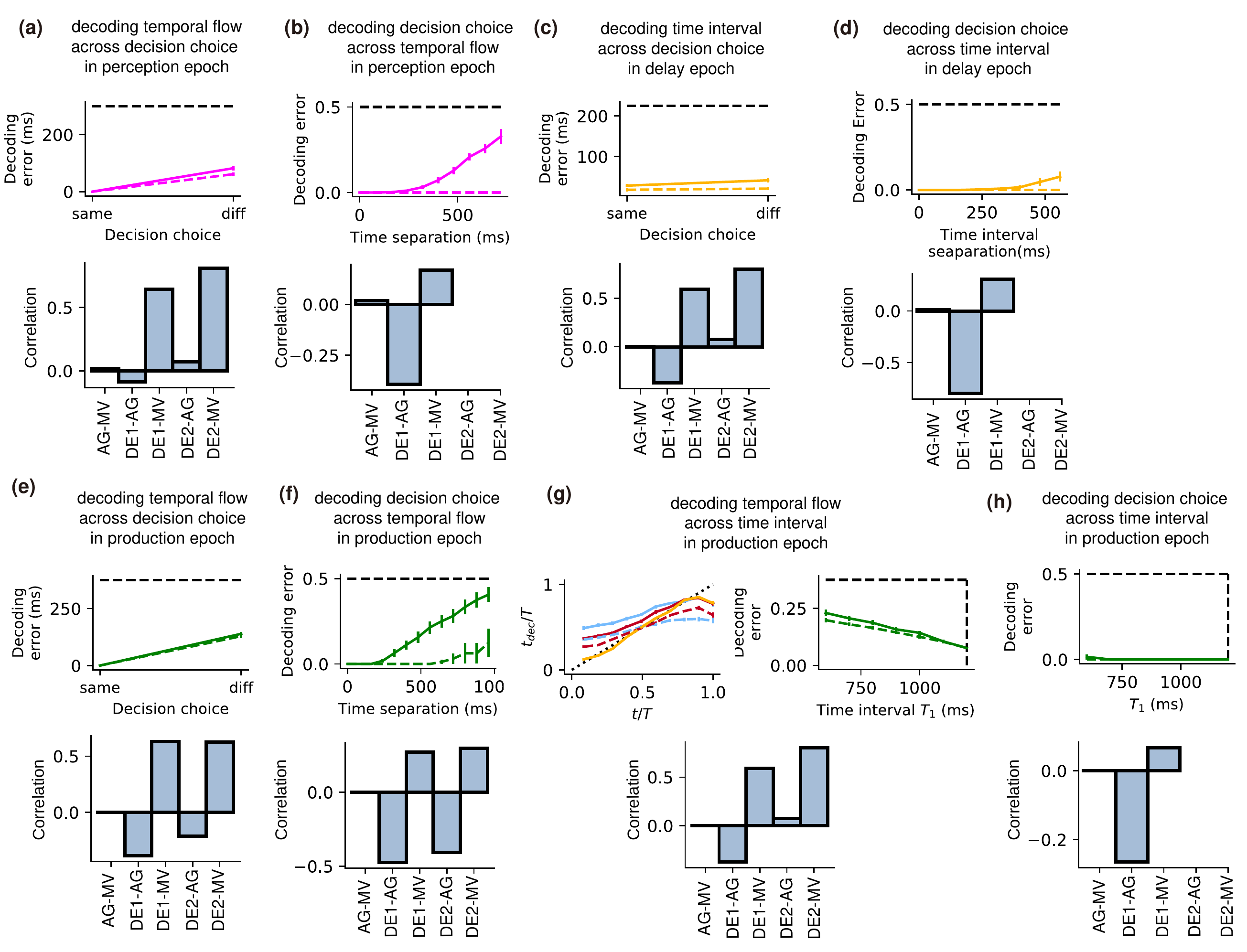}

\protect\protect\protect\caption{\textbf{\label{fig:Timed-decision-making-decoding}Decoding generalizability
in t-DM.} All panels are in the same style as \textbf{Fig. \ref{fig:Timed-spatial-reproduction-decoding}},
except that the non-temporal information in t-DM is the decision
choice. Note that in some panels (lower panels of \textbf{b}, \textbf{d},
\textbf{h}), the correlation between DE2 and AG as well as the correlation
between DE2 and MV are absent. The reason is that in these cases,
the decoding error is perfectly zero in all training configurations,
so the correlation is undefined. }
\end{figure}

\begin{figure}
\centering \includegraphics[scale=0.8]{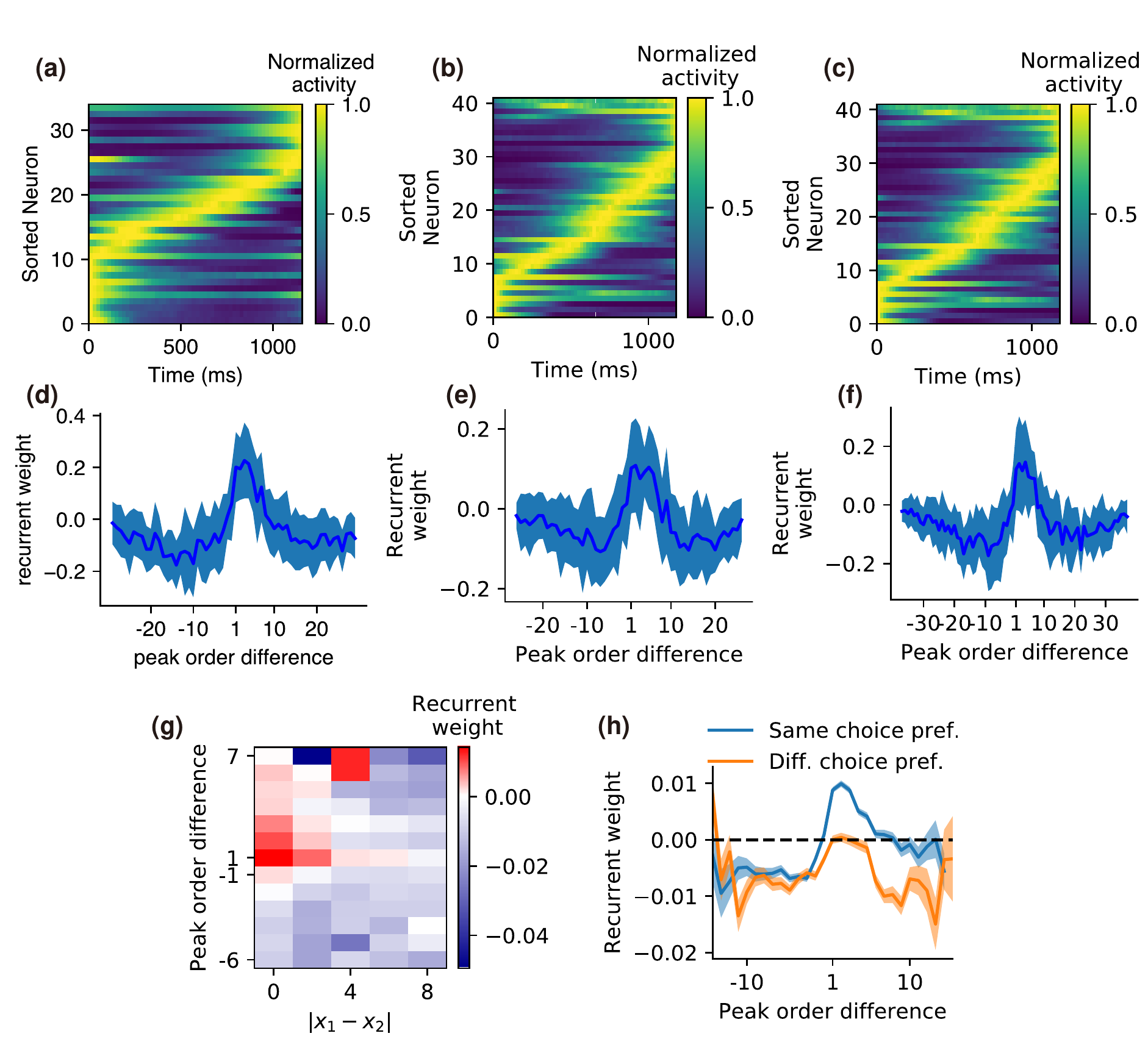}

\protect\protect\caption{\textbf{\label{fig:Sequential-activity-and-network-structure}Sequential
activity and network structure. }(\textbf{a}) The neuronal activity
(with maximum normalized to 1) in the production epoch of IP task
in an example training configuration, sorted according to peak time.
(\textbf{b, c}) The same as panel \textbf{a}, but for the stimulus1
(panel \textbf{b}) or stimulus2 (panel \textbf{c}) epoch of IC. (\textbf{d})
Mean (solid line) and s.d. (shaded belt) of the recurrent weights
as a function of the peak order difference between post- and pre-synaptic
neurons in the production epoch of IP. (\textbf{e, f}) The same as
panel \textbf{d}, but for the stimulus1 (panel \textbf{e}) or stimulus2
(panel \textbf{f}) epoch of IC. (\textbf{g}) Recurrent weight as
a function of the difference $|x_{1}-x_{2}|$ between the preferred
spatial locations of post- and pre-synaptic neurons and their peak
order difference in the production epoch of t-SR. (\textbf{h}) Recurrent
weight as a function of peak order difference in the sequence of
neurons with the same (blue) or different (orange) preferred decision
choices in the production epoch of t-DM. Shaded belt indicates s.e.m. }
\end{figure}

\begin{figure}
\centering \includegraphics[scale=0.65]{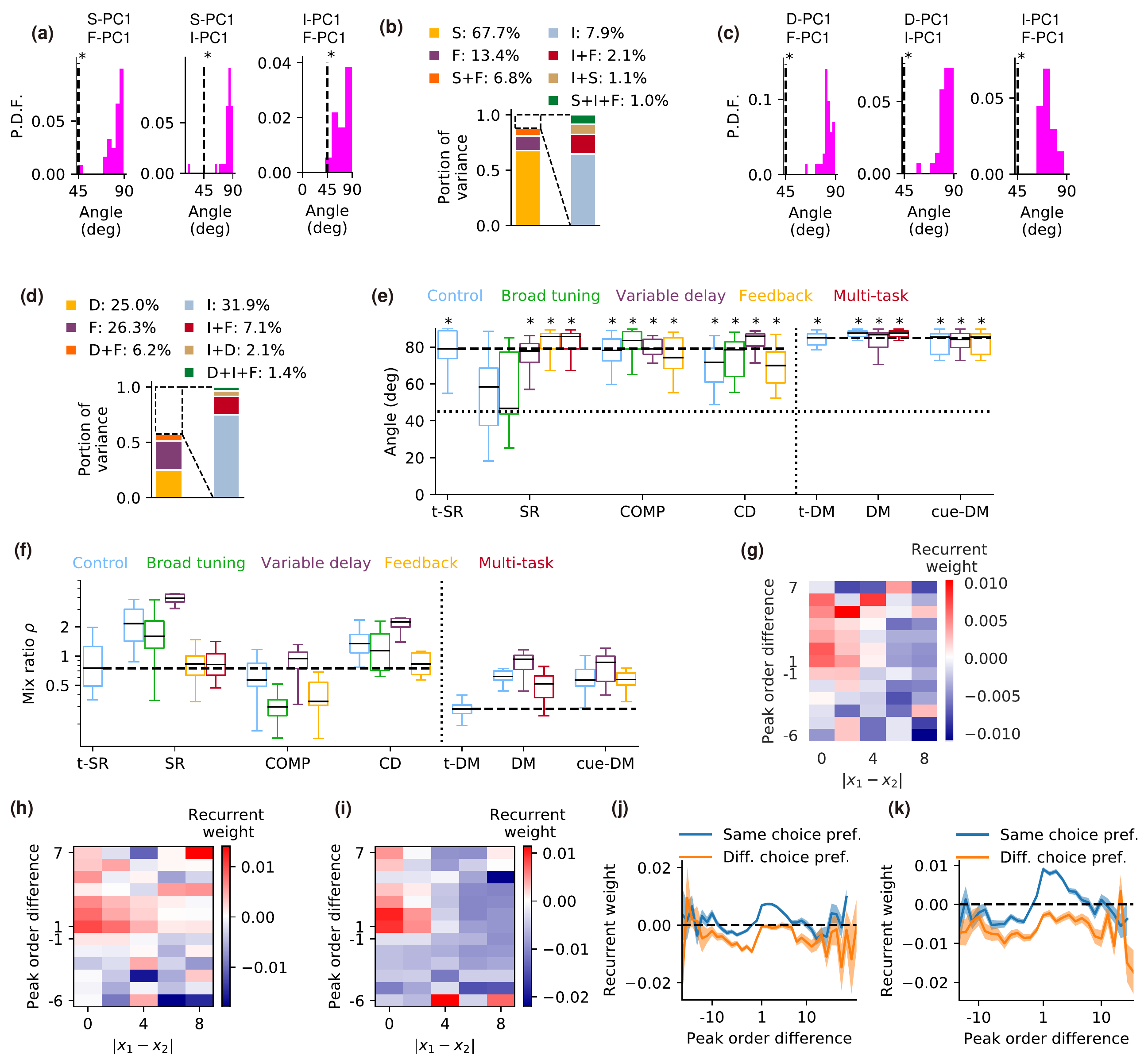}

\protect\protect\caption{\textbf{\label{fig:non-timing tasks}Coding geometry and network
structure in the absence of timing task requirement. }(\textbf{a,b})
The angle and mixed variance between the subspaces coding temporal
flow (F), time interval (I) and spatial information (S) in the delay
epoch of t-SR, in the same style as \textbf{Fig. 3i, j}. (\textbf{c,d})
Similar to panel \textbf{a,b}, except for the delay epoch of t-DM,
where the non-temporal information is decision choice (D). (\textbf{e})
The angle between the first temporal-flow PC and the first spatial
(in t-SR, SR, COMP and CD) or decision-choice (in t-DM, DM and
cue-DM) PC. Whisker plots: center line, median; box, 25th to 75th
percentiles; whiskers, $\pm1.5\times$ the interquartile range. In
t-SR and t-DM, the perception epoch is studied; in SR, COMP and
CD, the delay epoch is studied; in DM and cue-DM, the stimulus-presentation
epoch is studied. Asterisk indicates significant ($p<0.05$) larger
than $45^{\circ}$ (t test). The horizontal dotted line indicates
$45^{\circ}$, the vertical dotted line separates the spatial task
group (t-SR, SR, COMP and CD) from the decision-making task group
(t-DM, DM and cue-DM). The two horizontal dashed line indicate the
median values of t-SR and t-DM (which respectively are the only timing task in
each group) separately. (\textbf{f}) Mixed ratio $\rho$ in several
tasks, where $\rho=v_{\text{min}}/\min(v_{\text{time}},v_{\text{non-time}})$,
where $v_{\text{min}}$ is the mixed variance, $v_{\text{time}}$
and $v_{\text{non-time}}$ are the variance explained by temporal
and non-temporal information separately. (\textbf{g}) Recurrent weight
as a function of the difference $|x_{1}-x_{2}|$ between the preferred
spatial locations of post- and pre-synaptic neurons and their peak
order difference in the delay epoch of SR. (\textbf{h}) The same
as panel \textbf{g}, except for COMP. (\textbf{i}) The same as panel
\textbf{g}, except for CD. (\textbf{j}) Recurrent weight as a function
of peak order difference in the sequence of neurons with the same
(blue) or different (orange) preferred decision choices during the
presentation of the stimuli in cue-DM. Shaded belt indicates s.e.m.
(\textbf{k}) The same as panel \textbf{j}, except for DM.}
\end{figure}

\begin{figure}[tbph]
\centering \includegraphics[scale=0.65]{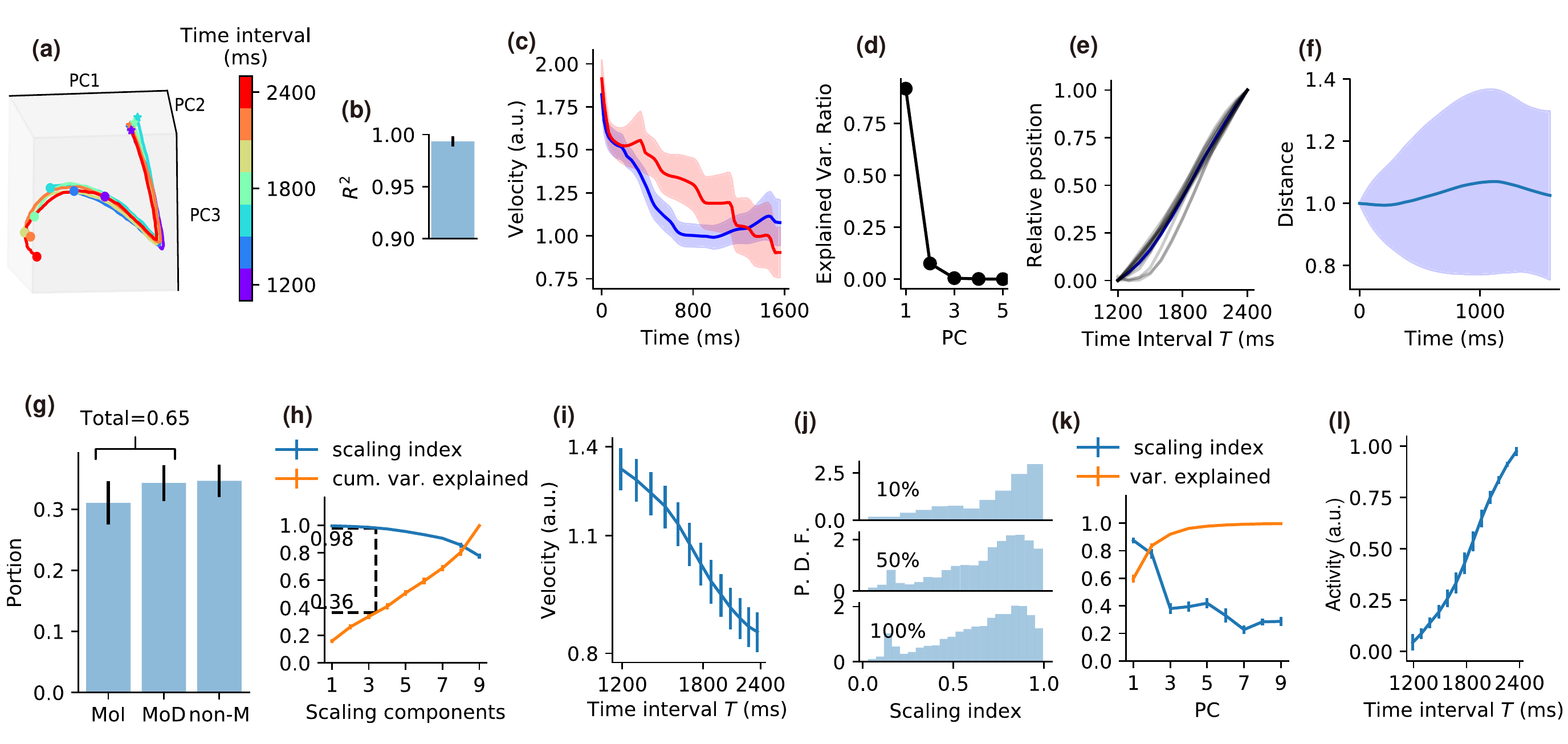}

\protect\protect\protect\caption{\textbf{\label{fig:IP-long-interval}Dynamics of the network when
trained to produce long time intervals.} (\textbf{a-b}) Perception
epoch. (\textbf{a}) Population activity in the perception epoch in
the subspace of the first three PCs. Colors indicate the time interval
$T$. Stars and circles respectively indicate the starting and ending
points of the perception epoch. (\textbf{b}) Coefficient of determination
($R^{2}$) that quantifies the overlap of the firing profiles of
individual neurons at different $T$s in the perception epoch, in
the same style as \textbf{Fig. 2d}. (\textbf{c-g}) Delay epoch (\textbf{c})
Trajectory speed in the delay epoch when $T=1200$ ms (blue) and
2400 ms (red), in the same style as \textbf{Fig. 2f}. (\textbf{d})
Ratio of explained variance of the first five PCs of manifold $\mathcal{M}$
at the end of the delay epoch, in the same style as \textbf{Fig.
2g}. (\textbf{e}) The position of the state at the end of the delay
epoch projected in the first PC of manifold $\mathcal{M}$ as a function
of $T$, in the same style as \textbf{Fig. 2h}. (\textbf{f}) The
distance between two adjacent curves in the delay epoch as a function
of time, in the same style as \textbf{Fig. 2i}. (\textbf{g}) The
portions of monotonically decreasing (MoD), monotonically increasing
(MoI) and non-monotonic (non-M) types of neurons tuned by $T$ at
the end of the delay epoch, in the same style as \textbf{Fig. 2k}.
(\textbf{h-l}) Production epoch. (\textbf{h}) Scaling index (blue)
and ratio of explained variance (orange) in the subspace spanned
by the accumulated scaling components, in the same style as \textbf{Fig.
2n}. (\textbf{i}) Trajectory speed in the subspace of the first three
scaling components, in the same style as \textbf{Fig. 2o}. (\textbf{j})
Probability distribution of the scaling indexes of single neurons,
in the same style as \textbf{Fig. \ref{fig:Interval-production-task.}b}.
(\textbf{k}) The scaling index and explained variance of principal
components, in the same style as \textbf{Fig. \ref{fig:Interval-production-task.}c}.
(\textbf{l}) Mean activity of the last scaling component as a function
of $T$, in the same style as \textbf{Fig. \ref{fig:Interval-production-task.}e}. }
\end{figure}


\bibliographystyle{ieeetr}
\bibliography{referenceSI}